\documentclass[fleqn,usenatbib]{mnras}

\usepackage{newtxtext,newtxmath}

\usepackage[T1]{fontenc}
\usepackage{ae,aecompl}

\usepackage{graphicx}	
\usepackage{amsmath}	
\usepackage{amssymb}	
\usepackage{bm}
\usepackage{adjustbox}

\usepackage{hyperref}

\def\prl{Phys.~Rev.~Lett.}

\def\apjl{ApjL}

\newcommand{\ez}{\hat{\bm{z}}}
\newcommand{\bh}{\bm{\hat{b}}}

\newcommand{\pc}{p_c}
\newcommand{\pg}{p_g}

\newcommand{\vrm}{{\rm v}}
\newcommand{\bvrm}{\bm{{\rm v}}}

\newcommand{\vadp}{{\rm v}_{{\rm A}, \Delta p}}
\newcommand{\bvadp}{\bm{{\rm v}_{{\rm A}, \Delta p}}}
\newcommand{\nub}{\nu_{\rm B}}
\newcommand{\imw}{{\rm Im}(\omega)}

\title[CR Acoustic Instabilities in Dilute Plasmas]{Sound-Wave Instabilities in Dilute Plasmas with Cosmic Rays: Implications for Cosmic-Ray Confinement and the Perseus X-ray  Ripples}

\author[Kempski, Quataert \& Squire]{
Philipp Kempski,$^{1,2}$\thanks{E-mail: philipp.kempski@berkeley.edu}
Eliot Quataert,$^{1}$
Jonathan Squire$^{3}$
\\
$^{1}$Department of Astronomy and Theoretical Astrophysics Center, University of California, Berkeley, CA 94720, USA \\
$^{2}$Kavli Institute for Theoretical Physics, University of California, Santa Barbara, CA 93106, USA \\
$^{3}$Department of Physics, University of Otago, 730 Cumberland St, North Dunedin, Dunedin 9016, New Zealand}
\date{Accepted XXX. Received YYY; in original form ZZZ}

\pubyear{2019}

\begin{document}
\label{firstpage}
\pagerange{\pageref{firstpage}--\pageref{lastpage}}
\maketitle

\begin{abstract}
Weakly collisional, magnetised plasmas characterised by anisotropic viscosity and conduction are ubiquitous in galaxies, halos and the intracluster medium (ICM). Cosmic rays (CRs) play an important role in these environments as well, by providing additional pressure and heating to the thermal plasma. We carry out a linear stability analysis of weakly collisional plasmas with cosmic rays using Braginskii MHD for the thermal gas. We assume that the CRs stream at the Alfv\'en speed, which in a weakly collisional plasma depends on the pressure anisotropy ($\Delta p$) of the thermal plasma. We find that this $\Delta p$-dependence introduces a phase shift between the CR-pressure and gas-density fluctuations. This drives a fast-growing acoustic instability: CRs offset the damping of acoustic waves by anisotropic viscosity and give rise to wave growth when the ratio of CR pressure to gas pressure is $\gtrsim \alpha \beta^{-1/2}$, where $\beta$ is the ratio of thermal to magnetic pressure, and $\alpha$, typically  $\lesssim 1$, depends on other dimensionless parameters. In high-$\beta$ environments like the ICM, this condition is satisfied for small CR pressures. We speculate that the instability studied here may contribute to the scattering of high-energy CRs and to the excitation of sound waves in galaxy-halo, group and cluster plasmas, including the long-wavelength X-ray fluctuations in \textit{Chandra} observations of the Perseus cluster. It may also be important in the vicinity of shocks in dilute plasmas (e.g., cluster virial shocks or galactic wind termination shocks), where the CR pressure is locally enhanced.     
\end{abstract}


\section{Introduction} \label{sec:intro}
The interstellar medium (ISM), the intracluster medium (ICM), and the halos of galaxy groups and Milky-Way-like galaxies are filled with hot and dilute gas, in which the electron/ion mean free paths along the magnetic field greatly exceed the particle gyroradii. Under such conditions, transport of heat and momentum is anisotropic and happens preferentially in the direction of the local magnetic field. The particle mean free path in these tenuous plasma environments can be large (i.e. the plasma is weakly collisional). As a result, anisotropic transport is efficient and can significantly affect the thermal and dynamical evolution of the gas.

The importance of  anisotropic conduction and viscosity in cluster environments has been underpinned by a variety of analytic theory and simulations. Anisotropic transport is an efficient driver of buoyancy instabilities (\citealt{mti_balbus}; \citealt{hbi}; \citealt{kunz12}) and significantly affects the gas dynamics in cluster simulations (e.g., \citealt{ruszkowski2010}; \citealt{parrish2012}; \citealt{yang16}; \citealt{barnes2019}; \citealt{kingsland19}).     

The ISM, galaxy halos, groups and the ICM are also permeated by a non-thermal population of relativistic particles known as cosmic rays (CRs). Even though they essentially propagate at the speed of light, their lifetime in galactic discs and halos can be quite long due to scattering off electromagnetic fluctuations. The waves responsible for the scattering can be Alfv\'en waves generated by the cosmic rays themselves through the streaming instability (\citealt{kp69}). Pitch-angle scattering by the excited waves isotropises the cosmic rays in the frame of the Alfv\'en waves. In this so-called self-confinement picture, cosmic rays are scattered towards isotropy in the Alfv\'en frame and collectively drift down their pressure gradient at the  Alfv\'en speed, provided that the pitch-angle scattering is sufficiently rapid. For slower pitch-angle scattering rates, the CR transport deviates from pure streaming at the Alfv\'en speed, but its exact nature remains uncertain (cosmic rays are believed to either diffuse or stream at super-Alfv\'enic speeds, or both; \citealt{skilling71}, \citealt{wiener2013}; \citealt{amato_blasi_18}). The self-confinement picture is in contrast to the extrinsic turbulence picture, where CRs are scattered primarily by extrinsic fluctuations that are not excited by the particles themselves. In this case, cosmic rays generally do not stream at Alfv\'enic speeds, even in the limit of fast scattering. In this work, we focus on self-confined cosmic rays.

The additional pressure force ($-\bm{\nabla}p_c$) and gas heating ($-\bm{{\rm v_{A}} \cdot \nabla}p_c$; \citealt{wentzel1971}) provided by the cosmic rays can be important for the dynamics and thermal evolution of gas in galaxies, halos and clusters (e.g., \citealt{bmv91}; \citealt{loew91}; \citealt{everett08}; \citealt{socrates08}; \citealt{guo08}; \citealt{zweibel_micro}; \citealt{ruszkowski17}; \citealt{zweibel17}; \citealt{jp_1}; \citealt{jp_2}; \citealt{ehlert18}; \citealt{farber18}; \citealt{kq19}). Cosmic rays can also directly affect MHD waves. For example, \cite{begelman_zweibel_1994}, hereafter BZ94, showed that CR heating can drive an acoustic instability in low-$\beta$ plasmas ($\beta \lesssim 1$).


The purpose of this work is to study sound waves in the presence of cosmic rays in magnetised, weakly collisional plasmas, i.e. plasmas with large anisotropic viscosity and conduction. We use the Braginskii MHD  closure for weakly collisional plasmas (\citealt{br65}) with anisotropic conduction and anisotropic pressure (the latter acts as an anisotropic viscosity), coupled to a 1-moment fluid equation for the cosmic-ray pressure. The cosmic rays are assumed to stream at the Alfv\'en speed $\vadp$, which in a weakly collisional plasma depends on the pressure anisotropy of the thermal plasma, $\Delta p$. We find that this dependence of $\vadp$ on $\Delta p$, which is not present in standard high-collisionality MHD, gives rise to a rapidly growing acoustic instability  (i.e. instability of the fast magnetosonic mode). The instability is driven by a phase shift between the CR pressure and gas density. Unlike the acoustic instability in BZ94, the Cosmic Ray Acoustic Braginskii (CRAB) instability that we find here is not driven by CR heating and does not require low $\beta$. In fact, the CRAB instability exists even at small CR pressures and has faster growth rates in high-$\beta$ systems. It is thus likely important in the ICM, in galactic halos and in the hot ISM.     

The remainder of this work is organised as follows. We present the CR--gas equations, examine the validity of our model and introduce characteristic timescales in Section \ref{sec:equations}.  We describe the CR-driven acoustic instability in Section \ref{sec:fast} and consider possible astrophysical  implications in Section \ref{sec:applications}. In Section \ref{sec:perseus} we speculate on the potential connection between the acoustic instability and the X-ray surface-brightness fluctuations observed in galaxy clusters such as Perseus (\citealt{fabian03}). In \ref{sec:shock} we hypothesise that the instability is likely important close to shocks, including the vicinity of the virial radius.  We discuss the potential contribution of the long-wavelength waves generated by the instability to the scattering of high-energy cosmic rays in Section \ref{sec:confine}.  We summarise our results in Section \ref{sec:conclusions}.

\section{Equations}\label{sec:equations}
We model the dilute, weakly-collisional plasmas filled with cosmic rays by using the Braginskii MHD equations coupled to a cosmic-ray pressure,
\begin{equation} \label{eq:cont}
\frac{\partial \rho}{\partial t} +  \bm{\nabla \cdot} ( \rho \bvrm ) = 0
\end{equation}
\begin{equation} \label{eq:mom}
\rho \frac{d \bvrm}{d t} = - \bm{\nabla} \Big( p_\perp + p_c + \frac{B^2}{ 8 \pi} \Big) + \frac{\bm{B \cdot \nabla B}}{4 \pi} + \bm{\nabla \cdot } \big( \bm{\hat{b} \hat{b}} \Delta p \big)
\end{equation}
\begin{equation} \label{eq:induction}
\frac{\partial \bm{B}}{\partial t} = \bm{\nabla \times} (\bm{\bvrm \times B})
\end{equation}
\begin{equation}\label{eq:s}
\rho T \frac{ds}{dt} = -\bm{\vadp \cdot \nabla}p_c + \mathcal{H} - \mathcal{C} - \bm{\nabla \cdot } \Big( \bm{\Pi \cdot \vrm} \Big) - \bm{\nabla \cdot Q}
\end{equation}
\begin{equation} \label{eq:pc}
\frac{dp_c}{dt} = -\frac{4}{3}p_c \bm{\nabla \cdot} ( \bvrm + \bvadp) - \bm{\vadp\cdot \nabla}p_c + \bm{\nabla \cdot} \big(\kappa \bm{\hat{b}\hat{b} \cdot \nabla} p_c \big),
\end{equation}
where  $\bvrm$ is the gas velocity, $\rho$ is the gas density, $\pg$ and $\pc$ are the gas and CR pressures respectively, $\bm{B}$ is the magnetic field (with unit vector $\bh$),  and $s=k_{\rm B} \ln(p_g/\rho^{\gamma}) / (\gamma -1)m_{\rm H}$ is the gas entropy per unit mass. $d / dt \equiv \partial / \partial t + \bm{{\rm v} \cdot \nabla}$ denotes a total (Lagrangian) time derivative. $\mathcal{H}$ and $\mathcal{C}$ are arbitrary volumetric heating and cooling rates. The Braginskii MHD pressure anisotropy (with viscosity $\nub$) is 
\begin{equation} \label{eq:Deltap}
    \Delta p = p_\perp - p_\parallel = 3 \rho \nub \big( \bm{\hat{b} \hat{b} : \nabla \vrm} - \frac{1}{3} \bm{\nabla \cdot \vrm} \big) = 3 \rho \nub \frac{d}{dt} \ln \frac{B}{\rho^{2/3}},
\end{equation}
where $\perp$ and $\parallel$ denote the directions perpendicular and parallel to the magnetic field (\citealt{br65}). $p_\perp$ and $p_\parallel$ are related to the total thermal pressure by
\begin{equation} \label{eq:pperp}
    p_\perp = p_g + \frac{1}{3} \Delta p.
\end{equation}
The viscous stress tensor in the gas-entropy equation depends on the pressure anisotropy and is given by
\begin{equation}
    \bm{\Pi} = - \Delta p \Big(\bm{\hat{b}\hat{b} - \frac{\mathcal{I}}{3}} \Big).
\end{equation}
Note that in the absence of background flow, the perturbed $\bm{\nabla \cdot} (\bm{\Pi \cdot \vrm})$ in the gas-entropy equation is second-order and does not contribute in our linear analysis. $\bm{Q}$ in equation \ref{eq:s} is the anisotropic thermal heat flux, 
\begin{equation}
    \bm{Q} = - \kappa_{\rm B} \bm{\hat{b}\hat{b}\cdot \nabla}T,
\end{equation}
where $\kappa_{\rm B}$ is the thermal conductivity.\footnote{While in this work we assume that the heat transport is diffusive, we note that recent particle-in-cell simulations suggest that the transport may also occur down the temperature gradient at the whistler phase speed (\citealt{rob-clark18}).}
$\vadp$ in equations \ref{eq:s} and \ref{eq:pc} is the Alfv\'en speed in the presence of pressure anisotropy,
\begin{equation} \label{eq:va_mod}
    \bvadp = \frac{\bm{B}}{\sqrt{4 \pi \rho}} \Big(1 + \frac{4\pi \Delta p}{B^2} \Big).
\end{equation}

 We assume that cosmic rays stream down their pressure gradient at the Alfv\'en velocity $\bvadp$ and we also include CR diffusion along the magnetic field, for which we assume a constant diffusion coefficient $\kappa$. We note that formally CRs stream with velocity $\bm{{\rm v}_{\rm st}} = -{\rm sgn}(\bm{\hat{b} \cdot \nabla}p_c) \bvadp$. This ensures that cosmic rays stream along the magnetic field down their pressure gradient and makes the CR heating term $-\bm{{\rm v}_{\rm st} \cdot \nabla}p_c$ positive definite. In our linear stability analysis cosmic rays stream at $\bvadp$, as we consider background equilibria which satisfy $-\bm{\vadp \cdot \nabla} p_c >0$.\footnote{We do not explicitly include background gradients in our linear stability calculation, and so we treat the background as effectively uniform. However, for $\nabla p_c$ to have a well-defined sign in the linear stability calculation, so that $- \bm{\vadp \cdot \nabla} p_c$ is positive definite, $p_{c}$ cannot be exactly uniform. A background CR pressure gradient is necessary. However, we can neglect terms associated with  background gradients as long as the additional timescale introduced by a spatially varying background is much longer than the timescale associated with the acoustic instability considered here. This is well-motivated given the fast growth rates of the instability, which can be comparable to the sound oscillation frequency. We can, for example, consider an equilibrium with $-\bm{\vadp \cdot \nabla} p_c =   \mathcal{C}$, where $\mathcal{C}$ is a cooling rate with an associated cooling frequency $\omega_c$. Our linear stability calculation can neglect background gradients provided that  $\omega_s, \imw \gg  \omega_c $. If this is satisfied,  the equlibrium $\rho$, $\pg$ and $\pc$ can be treated as uniform without significantly changing the results. \label{ftn:pos_def}} 

\subsection{MHD Waves in Weakly Collisional Plasmas} \label{sec:damping}
In this section we ignore cosmic rays and review how standard MHD waves are modified at low collisionality. The pressure anisotropy changes the Alfv\'en speed (eq. \ref{eq:va_mod}) because it modifies the effective magnetic tension, as can be seen by rewriting eq. \ref{eq:mom}:
\begin{equation} \label{eq:mom2}
\rho \frac{d \bvrm}{d t} = - \bm{\nabla} (p_\perp + p_c + \frac{B^2}{ 8 \pi}) + \bm{\nabla \cdot} \Big(\frac{\bm{BB}}{4\pi} \big( 1 + \frac{4\pi \Delta p}{B^2} \big) \Big).
\end{equation}
Note that the factor $\big( 1 + 4\pi \Delta p/B^2 \big)$ enters the effective magnetic tension term (which is responsible for Alfv\'en waves). The dispersion relation for Alfv\'en waves can then be easily derived by assuming wave perturbations $\propto \exp \Big(i \bm{k \cdot r} - i \omega t \Big)$, crossing the momentum equation twice with $\bm{k}$ and noting that $\delta \Delta p = 0$ for Alfv\'enic perturbations (which are incompressible and do not change the $B$-field strength, see eq. \ref{eq:Deltap}). From this, equation \ref{eq:va_mod} follows.

The pressure anisotropy has a different effect on the slow and fast modes, which are viscously damped in Braginskii MHD (still ignoring cosmic rays). By inserting eq. \ref{eq:Deltap} into eq. \ref{eq:mom} (and noting eq. \ref{eq:pperp}), we obtain:
\begin{equation} \label{eq:mom3}
\rho \frac{d \bvrm}{d t}  = ... + \bm{\nabla \cdot}  \Big(3 \rho \nub \big(\bm{\hat{b}\hat{b}} - \frac{\mathcal{I}}{3} \big) \big( \bm{\hat{b} \hat{b} : \nabla \vrm } - \frac{1}{3} \bm{\nabla \cdot \vrm} \big)  \Big).  
\end{equation}
This diffusion operator associated with the Braginskii viscosity damps the fast and slow magnetosonic waves, because they involve perturbations that linearly generate $\delta \Delta p$, unlike the linearly undamped Alfv\'en waves for which $\delta \Delta p=0$. In the weak damping limit, the fast and slow modes are damped at a rate (\citealt{br65}; \citealt{parrish2012}):
\begin{equation} \label{eq:damping}
    \omega_\nu = \frac{\nub k^2}{6} \Big( (\bm{\hat{k} \cdot \hat{\vrm}}) - 3 (\bm{\hat{b} \cdot \hat{k}})(\bm{\hat{b} \cdot \hat{\vrm}})  \Big)^2,
\end{equation}
where $\bm{\hat{\vrm}}$ is the unit vector in the direction of the mode's perturbed velocity. We will show that in the presence of cosmic rays this is strongly modified, and sound waves can instead be linearly unstable.

\subsection{Validity of the Model} \label{sec:validity}
 Our CR--Braginskii MHD fluid model requires that the CR scattering rate is fast, so that eq. \ref{eq:pc} appropriately describes the CR pressure evolution. It also requires that the collision time of the thermal ions is short compared to the macroscopic timescales of interest (so that a weakly collisional, rather than collisionless, treatment is appropriate for the thermal plasma). In what follows, we check the validity of the CR--Braginskii MHD fluid model, focusing on the ICM and the hot phase of the ISM. 

For the CR pressure equation (eq. \ref{eq:pc}) to be a good model of the cosmic rays, the GeV CR collision frequency must be large. It is the GeV CRs that are important, as they dominate the bulk CR energy. The CR collision frequency is the rate at which the pitch angle changes by order unity, due to scattering by EM fluctuations of magnitude $\delta B_\perp$ at the resonant wavelength:
\begin{equation} \label{eq:cr_nu}
\nu_{\rm CR} \sim \Omega \Big( \frac{\delta B_\perp}{B} \Big)^2 \sim  10^{-8} \ {\rm s} ^{-1}  \  \Big(\frac{\gamma_{c}}{1}\Big)^{-1} \frac{B}{1 \ {\rm \mu G}}  \Big( \frac{\delta B_\perp / B}{10^{-3}} \Big)^2,
\end{equation}
where $\gamma_c$ is the CR Lorentz factor and $\delta B_\perp$ is evaluated for fluctuations whose wavelength parallel to the mean B-field is of order the Larmor radius of the GeV particles. Models of CR observations in the Milky Way  based on pure diffusion infer a CR diffusion coefficient $\kappa \sim 10^{28} - 10^{29} \ {\rm cm^2 \ s^{-1}}$ depending on assumptions about the CR halo size (e.g., \citealt{Linden2010}). This motivates the choice of $\delta B_\perp /B \sim 10^{-3}$ used in \eqref{eq:cr_nu}, as $\delta B_\perp /B \sim 10^{-3}$ corresponds to a GeV CR diffusion coefficient $\kappa \sim c^2 / \nu_{\rm CR} \sim 10^{29} \ {\rm cm^2 \ s^{-1}}$ in a $1 \ {\rm \mu G}$ field. However, this observationally inferred CR diffusion coefficient is not necessarily appropriate if CR streaming is the dominant transport process, as is theoretically favoured for the low-energy CRs that dominate the total energy density (these low-energy cosmic rays are the most likely to be adequately described by the fluid model used in this paper; \citealt{blasi12}). In the case of streaming transport the diffusion coefficient may be $\ll 10^{28}-10^{29} \ {\rm cm^2 \ s^{-1}}$. In particular, in the hot ISM and ICM damping processes are weaker than in the cold/neutral ISM and so the streaming instability can grow to large amplitudes (e.g., Figure 1 in \citealt{amato_blasi_18}).  For example, if $\delta B_\perp / B \sim 10^{-2}$ then $\nu_{\rm CR} \sim 10^{-6} \ {\rm s^{-1}}$ and the CR diffusion coefficient is significantly smaller,  $\kappa \sim 10^{27} \ {\rm cm^2 \ s^{-1}}$.

The weakly collisional fluid model used in this paper requires that the ion-ion collision frequency is larger than the rate of change of all fields, $\omega \lesssim  \nu_{\rm ii}$. We note that $\omega \ll \nu_{\rm ii}$ is formally required in deriving the equations of Braginskii MHD. We will consider $\omega_s \lesssim \nu_{\rm ii}$ in our calculations (where $\omega_s$ is the adiabatic sound frequency and typically the largest frequency in the problem), but our main conclusions do not change if we choose a smaller upper limit on $\omega_s$. We now separately estimate the ion collision rates in the ICM and the hot ISM. We will show that the CR scattering rate is much higher than the ion-ion collision frequency in both the ICM and hot ISM.

\subsubsection{ICM \label{sec:icm}}
Under typical ICM conditions, the plasma is magnetised and the collisionality is low. For representative ICM temperatures and densities, the ion-ion collision frequency is
\begin{equation}
\nu_{\rm ii} \sim \frac{  n_{\rm i} e^4 \pi \ln \Lambda }{m_{\rm i}^{1/2} (k_{\rm B} T)^{3/2}} \sim 8 \times 10^{-14} \ {\rm s} ^{-1} \ \Big(\frac{T}{5 \times 10^7 \ \rm{K}} \Big)^{-3/2} \frac{n_{\rm i}}{0.01 \ {\rm cm}^{-3}},
\end{equation}
for a Coulomb logarithm $\ln \Lambda \approx 38$. This corresponds to a collision time of approximately 0.4 Myrs. We note that $\nu_{\rm CR} \gg \nu_{\rm ii}$ (see eq. \ref{eq:cr_nu}) and so the CR--Braginskii MHD fluid model is a  good description for the ICM if we consider fields that vary at a frequency $\omega \ll \nu_{\rm ii}$ (see Section \ref{sec:params} for how this translates into constraints on the characteristic frequencies in our problem).  

\subsubsection{Hot ISM} \label{sec:hotism}
The plasma filling the hot ISM is cooler, so that the ion-ion collision frequency is larger than in the ICM:
\begin{equation}
\nu_{\rm ii} \sim 3 \times 10^{-11} \ {\rm s} ^{-1} \ \Big(\frac{T}{  10^6 \ \rm{K}} \Big)^{-3/2} \frac{n_{\rm i}}{0.01 \ {\rm cm}^{-3}}
\end{equation}
(for $\ln \Lambda \approx 32$). This ion-ion collision frequency is still, however, significantly smaller than the CR collision frequency (order unity pitch angle change; see eq. \ref{eq:cr_nu}). Just like in the ICM,  the CR--Braginskii MHD formulation is therefore well motivated in the hot ISM as long as we consider $\omega \ll \nu_{\rm ii}$. 


\subsection{Dimensionless Parameters and Characteristic Timescales} \label{sec:params}
We define the ratio of CR pressure to gas pressure,
\begin{equation} \label{eq:eta}
 \eta \equiv \frac{p_c}{p_g}   ,
\end{equation}
and the ratio of thermal to magnetic pressure, 
\begin{equation} \label{eq:beta}
    \beta \equiv \frac{8\pi p_g} {B^2} .
\end{equation}
The key frequencies in this problem are the gas sound frequency (with $c_s$ being the adiabatic gas sound speed),
\begin{equation} \label{eq:ws}
    \omega_s \equiv k c_s ;
\end{equation}
the Alfv\'en \textit{and} CR-heating frequency,
\begin{equation}\label{eq:wa}
\omega_a \equiv  \bm{k \cdot \vrm_{\rm A}};
\end{equation}
the cosmic-ray diffusion frequency,
\begin{equation}\label{eq:wd}
\omega_d \equiv \kappa \  (\bm{\hat{b} \cdot k})^2 ;
\end{equation}
the Braginskii viscous frequency,
\begin{equation}
    \omega_{\rm B} \equiv \nub (\bm{\hat{b}\cdot k})^2 \approx \frac{p_g}{3 \rho \nu_{\rm ii}} (\bm{\hat{b}\cdot k})^2;
\end{equation}
and the conductive frequency
\begin{equation}
    \omega_{\rm cond} \equiv \chi_{\rm B} (\bm{\hat{b}\cdot k})^2,
\end{equation}
where $\chi_{\rm B} = \kappa_{\rm B} / n k_{\rm B}$ is the thermal diffusion coefficient. We define the Braginskii viscous scale,
\begin{equation} \label{eq:lnub}
    l_{\nub} \equiv \frac{\nub}{c_s} \sim l_{\rm mfp} ,
\end{equation}
where $l_{\rm mfp}$ in the last step is the ion mean free path.
We can relate the diffusive timescales by defining the thermal Prandtl number,
\begin{equation} \label{eq:Pr}
    {\rm Pr} \equiv \frac{\nub}{\chi_{\rm B}},
\end{equation}
and the ratio of the CR diffusion coefficient to the Braginskii viscosity,
\begin{equation} \label{eq:phi}
    \Phi \equiv \frac{\kappa}{\nub}.
\end{equation}
It is commonly assumed that the heat flow is dominated by electrons, such that for a typical plasma ${\rm Pr} \sim 10^{-2}$ (set by the ion-to-electron mass ratio). This assumption is, however, not well motivated when the timescales of interest are shorter than the ion-electron temperature equilibration time (which is longer than the ion-ion collision time by a square root of the ion to electron mass ratio). This is the case in this work, where we consider sound waves at low collisionalities. A more accurate calculation should therefore consist of two entropy equations and two heat fluxes, one for each species. We avoid this complication in the main text of this paper by considering a single heat flux with varying conductivity: $\omega_{\rm cond} = \omega_{\rm B}$ (${\rm Pr = 1}$, $\sim$ heat flux carried by ions) and $\omega_{\rm cond} = 100 \omega_{\rm B}$ (${\rm Pr = 0.01}$, $\sim$ heat flux carried by electrons). We show in Appendix \ref{app:2F} and Figure \ref{fig:twoF} that our conclusions do not change when a two-fluid electron-ion system is considered instead, and that ${\rm Pr} \sim 1$ is a somewhat better approximation to the two-fluid results (a similar two-fluid electron-ion system was used in the context of cluster sound waves by \citealt{zweibel2018}).

$\Phi$ in eq. \ref{eq:phi} relates the Braginskii viscous frequency to the CR diffusion frequency,
\begin{equation}
     \Phi = \frac{\omega_d}{\omega_{\rm B}}  \sim \frac{c^2}{c_s^2} \frac{\nu_{\rm ii}}{\Omega (\delta B_\perp / B)^2},
\end{equation}
where $c$ is the speed of light. For typical ICM parameters,
\begin{equation}
    \Phi \sim 2 \ \Big(\frac{T}{5 \times 10^7 \  \rm{K}} \Big)^{-5/2} \frac{n_{\rm i}}{0.01 \ {\rm cm}^{-3}}  \Big( \frac{B}{1 \ {\rm \mu G}}\Big)^{-1}  \Big( \frac{\delta B_\perp / B}{10^{-3}} \Big)^{-2}.
\end{equation}
This suggests that $\Phi \sim 1$ in the ICM  (or $\Phi \ll 1$, if $\delta B_\perp / B \gg 10^{-3}$). $\Phi \gg 1$ for typical temperatures in the hot ISM, unless $\delta B_\perp / B \gg 10^{-3}$ (which is plausible, see discussion in Section \ref{sec:validity}). Motivated by these results, we will focus primarily on $\Phi = 0$ ($\omega_d = 0$), $\Phi = 1$ ($\omega_d = \omega_{\rm B}$) and $\Phi = 10$ ($\omega_d = 10 \omega_{\rm B}$).


$\omega_s$ is the largest characteristic frequency in the $\beta > 1$ plasmas that we focus on. We require that $ \omega_s \lesssim \nu_{\rm ii}$ so that the weakly collisional description is appropriate  (see Section \ref{sec:validity}),  which translates into
\begin{equation}\label{eq:valid_slow}
    \frac{\omega_s}{\nu_{\rm ii}} \sim \frac{\omega_{\rm B}}{\omega_s} \lesssim 1.
\end{equation}

The ICM is of primary interest in this work and so we will focus mainly on the high-$\beta$ limit ($\beta \sim 100$ unless specified otherwise). 

\subsection{Linearised Equations}

 We consider a uniform and static background equilibrium with $\mathcal{H} = \mathcal{C}$, i.e. all background fluid variables are assumed to be spatially constant. Thus, there are no background gradients in the linear stability analysis (see the comment regarding the CR pressure gradient in Footnote \ref{ftn:pos_def}). Without loss of generality, we consider a vertical magnetic field, $\bm{B} = B \ez$.

We carry out a linear stability calculation of the CR--gas equations (see Section \ref{sec:equations}). All perturbed quantities are assumed to vary as $\delta X(\bm{r}, t) \propto \exp \Big(i \bm{k \cdot r} - i \omega t \Big)$. Without loss of generality, we take $\bm{k}$ in the $xz$-plane, $\bm{k} = k \sin \theta \bm{\hat{x}}  + k \cos \theta \bm{\hat{z}}$. Alfv\'en waves can be isolated as described in Section \ref{sec:damping}, which remains valid in the presence of CRs. The remaining modes can be found by considering all linearised equations excluding the $y$-component of the momentum and induction equations:
\begin{equation} \label{eq:drho}
    \omega \frac{\delta \rho}{\rho} = \bm{k \cdot {\rm v}},
\end{equation}
\begin{gather}
\begin{aligned}\label{eq:dxi_x}
    \omega {\rm v}_x = & k_x \frac{c_s^2}{\gamma}  \frac{\delta p_g}{p_g} - \omega_a {\rm v_A} \frac{\delta B_x}{B} + k_x {\rm v_A^2} \frac{\delta B_z}{B} +  \frac{2}{3}i \frac{k_x}{k_z} \omega_{\rm B} {\rm v}_z  \\ & - \frac{1}{3}i \frac{k_x^2}{k_z^2} \omega_{\rm B} {\rm v}_x + \eta k_x \frac{c_s^2}{\gamma} \frac{\delta p_c}{p_c} ,
\end{aligned}
\end{gather}
\begin{gather}
\begin{aligned}\label{eq:dxi_z}
    \omega {\rm v}_z = & k_z \frac{c_s^2}{\gamma}  \frac{\delta p_g}{p_g}   - \omega_a {\rm v_A} \frac{\delta B_z}{B} + k_z {\rm v_A^2} \frac{\delta B_z}{B}   -\frac{4}{3}i  \omega_{\rm B} {\rm v}_z  \\ & + \frac{2}{3}i \frac{k_x}{k_z} \omega_{\rm B} {\rm v}_x + \eta  k_z \frac{c_s^2}{\gamma} \frac{\delta p_c}{p_c},
\end{aligned}
\end{gather}
\begin{equation} \label{eq:dB_x}
    \omega \frac{\delta B_x}{B} = - k_z {\rm v}_x,
\end{equation}
\begin{equation} \label{eq:dB_z}
    \omega \frac{\delta B_z}{B} =  k_x {\rm v}_x,
\end{equation}
\begin{equation} \label{eq:dpg}
    \omega \frac{\delta p_g}{p_g} = \gamma \bm{k \cdot {\rm v}}  - i(\gamma-1) \omega_{\rm cond} \Big( \frac{\delta p_g}{p_g} - \frac{\delta \rho}{\rho} \Big) + \eta (\gamma -1) \omega_a \frac{\delta p_c}{p_c} ,
\end{equation}
\begin{gather}
\begin{aligned} \label{eq:dpc}
    \omega \frac{\delta p_c}{p_c} = & \frac{4}{3} \bm{k \cdot {\rm v}} - \frac{2}{3} \omega_a \frac{\delta \rho}{\rho}  + \frac{8}{3}i \frac{\omega_{\rm B}}{\omega_a} k_z {\rm v}_z - \frac{4}{3}i  \frac{\omega_{\rm B}}{\omega_a} k_x {\rm v}_x  \\ & + (\omega_a - i \omega_d) \frac{\delta p_c}{p_c},
\end{aligned}
\end{gather}
where $\gamma = 5/3$ is the gas adiabatic index. We find the exact eigenmodes by solving the full matrix eigenvalue problem using MATLAB.

\section{The Cosmic-Ray Acoustic Instability in Braginskii MHD} \label{sec:fast}

Before we show growth rates and simplified dispersion relations, we discuss the physical mechanism that drives the sound-wave instability.
\subsection{Driving Mechanism and Negative Effective Viscosity from Cosmic Rays} \label{sec:driving}

 \begin{figure}
  \centering
    \includegraphics[width=0.45\textwidth]{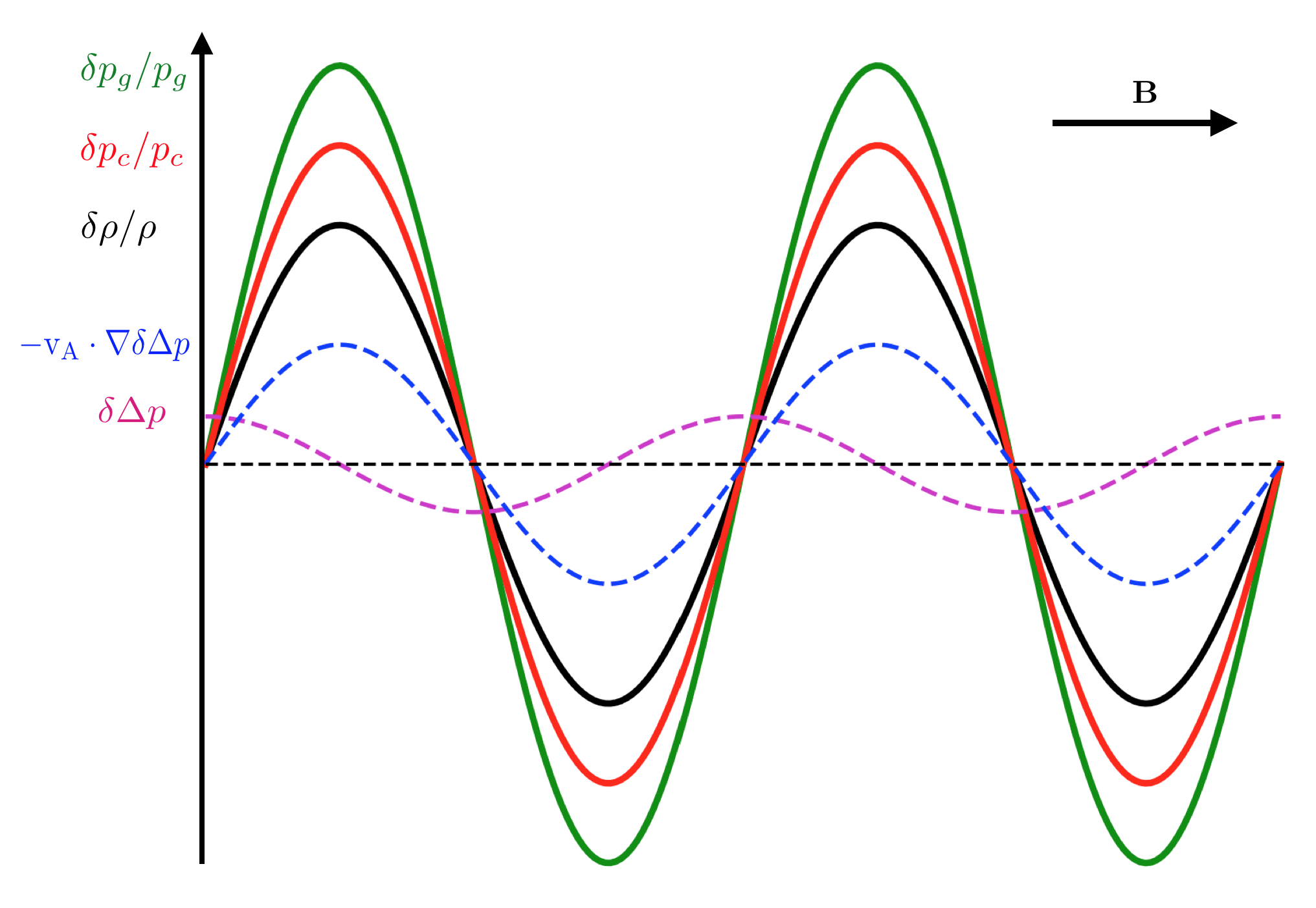}
  \caption{Schematic of the mechanism driving the acoustic instability. The solid waveforms show the leading-order adiabatic gas-density, gas-pressure and CR-pressure perturbations in the frame comoving with the sound wave at the phase speed ${\rm v_{ ph}}$ in the $\bm{B}$-direction. The pressure-anisotropy perturbation $\delta \Delta p \sim  \nub {\rm v_{ph}} d \delta \rho /dz$ (eq. \ref{eq:Deltap} in the moving frame; dashed magenta line) has a $90^\circ$ phase shift relative to $\delta \rho / \rho$, $\delta p_c / p_c$ and $\delta p_g / p_g$. Without CRs this phase shift leads to the well-known damping of acoustic waves by anisotropic viscosity. In the presence of cosmic rays, the work done by $\delta \Delta p$ on the CRs (dashed blue line and eq. \ref{eq:work_on_cr}) is positive in regions where $\delta p_c >0$: it therefore amplifies $\delta p_c$ in the regions where $\delta p_c >0$ and reduces $\delta p_c$ where $\delta p_c <0$. This drives the perturbations and leads to wave growth.   \label{fig:cartoon} }
\end{figure}

 The instability is driven by a phase shift between the CR-pressure and the gas-density perturbations, which comes from the dependence of the Alfv\'en speed on $\Delta p$ (see eq. \ref{eq:phase_shift} in Section \ref{sec:1D}). Such phase shifts generally occur in the presence of diffusion operators (e.g., CR diffusion also leads to a phase shift between $\delta p_c$ and $\delta \rho$). However, these tend to damp the perturbations instead of driving instabilities. The phase shift introduced by ${\rm v}_{{\rm A}, \Delta p}$ gives rise to an instability because it introduces an additional diffusion operator in the momentum equation (eq. \ref{eq:mom}) which can have a negative diffusivity (negative viscosity) and thus generate wave growth.\footnote{A negative diffusion coefficient can be thought of as standard diffusion reversed in time.}
 
 For standard, collisional MHD sound waves, the CR pressure responds essentially adiabatically to density fluctuations in the limit $\omega_a, \ \omega_d \ll \omega_s$ (otherwise the CR response is generally non-adiabatic, see eq. \ref{eq:dpc} with $\omega_{\rm B} = 0$). In weakly collisional plasmas the CR pressure also responds to changes in the pressure anisotropy, which in turn depends on the rate of change of $\delta \rho$. This phase shift (in addition to the adiabatic response) provides a driving force to the wave, which can win over the damping by anisotropic viscosity and give rise to instability.
 
 The key term for driving the instability is the compression work done on the cosmic rays by the pressure anisotropy,
 \begin{equation} \label{eq:work_on_cr}
     \frac{d \delta p_c}{dt} = - \frac{4 p_c}{3 \rho {\rm v_A^2}}  \bm{{\rm v_A} \cdot \nabla} \delta \Delta p + ... \ ,
 \end{equation}
 which comes from the $\bm{\nabla \cdot \vadp}$ term in equation \ref{eq:pc}. To see what this term does to the sound wave, it is useful to consider the frame comoving with the wave in the $\bm{B}$-direction. In this frame, moving at a phase speed $v_{\rm ph}$, the wave profile is stationary to leading order  (i.e. ignoring the growth or damping of the wave) and is shown in Figure \ref{fig:cartoon}. $\delta \Delta p \sim  \nub {\rm v_{ph}} d \delta \rho /dz$ (eq. \ref{eq:Deltap} in the moving frame) has a $90^\circ$ phase shift relative to $\delta \rho / \rho$, $\delta p_c / p_c$ and $\delta p_g / p_g$, and without cosmic rays this phase shift leads to wave damping. However, the work done by $\Delta p$ on the CRs (eq. \ref{eq:work_on_cr}) is positive in regions where $\delta p_c >0$, as shown in Figure \ref{fig:cartoon}. This process amplifies $\delta p_c$ in the regions where $\delta p_c >0$ and reduces $\delta p_c$ where $\delta p_c <0$. This drives the perturbations and leads to wave growth.  
 
This driving manifests itself mathematically as a negative effective diffusion coefficient (i.e. negative viscosity) introduced by the cosmic rays in the momentum equation. This can be demonstrated by inserting equation \ref{eq:pc} into equation \ref{eq:mom} and assuming wave perturbations proportional to $f(\bm{k \cdot r} - \omega t)$ propagating at the sound speed. Ignoring all other non-diffusive terms in the momentum equation, this gives:
\begin{gather}
\begin{aligned}
 \label{eq:mom4}
\rho \frac{d \bvrm}{d t} & = ... + \bm{\nabla \cdot}  \Big(3 \rho \nub \big(\bm{\hat{b}\hat{b}} - \frac{\mathcal{I}}{3} \big) \big( \bm{\hat{b} \hat{b} : \nabla \vrm} - \frac{1}{3} \bm{\nabla \cdot \vrm} \big)  \Big)   \\ & - \frac{4\eta \sqrt{\beta}}{3\sqrt{2 \gamma}}(\bm{\hat{b} \cdot \hat{k}})  \bm{\nabla} \Big( 3 \rho \nub \big( \bm{\hat{b} \hat{b} : \nabla \vrm} - \frac{1}{3} \bm{\nabla \cdot \vrm } \big) \Big).
\end{aligned}
\end{gather}
The first term is the damping by Braginskii viscosity, the second term is the additional diffusive term that comes from $\delta p_c$. Longitudinal acoustic waves approximately satisfy $\bvrm \parallel \bm{k}$. If $\bm{\hat{b} \cdot \hat{k}} > 0$, the last term acts as a diffusion operator with negative viscosity if $\bm{\hat{b} \hat{b} : \nabla \vrm} - \frac{1}{3} \bm{\nabla \cdot \vrm } \gtrsim 0$, i.e. $\cos^2 \theta \gtrsim 1/3$ ($\theta \lesssim 55^\circ$), where $\theta$ is the angle between $\bm{k}$ and $\bm{B}$. For $\cos^2 \theta \lesssim 1/3$ ($\theta \gtrsim 55^\circ$), it acts as a diffusion operator with negative viscosity for longitudinal acoustic modes propagating in the opposite direction, $\bm{\hat{b} \cdot \hat{k}} < 0$.\footnote{Note that the magnetic-field direction in $\bm{\hat{b} \cdot \hat{k}}$ in eq. \ref{eq:mom4} comes from the direction of CR streaming (which occurs in the $\bm{\hat{b}}$-direction due to our assumption that the background CR pressure decreases in the direction of the magnetic field).} 

\begin{figure*}
  \centering
  \begin{minipage}[b]{0.32\textwidth}
    \includegraphics[width=\textwidth]{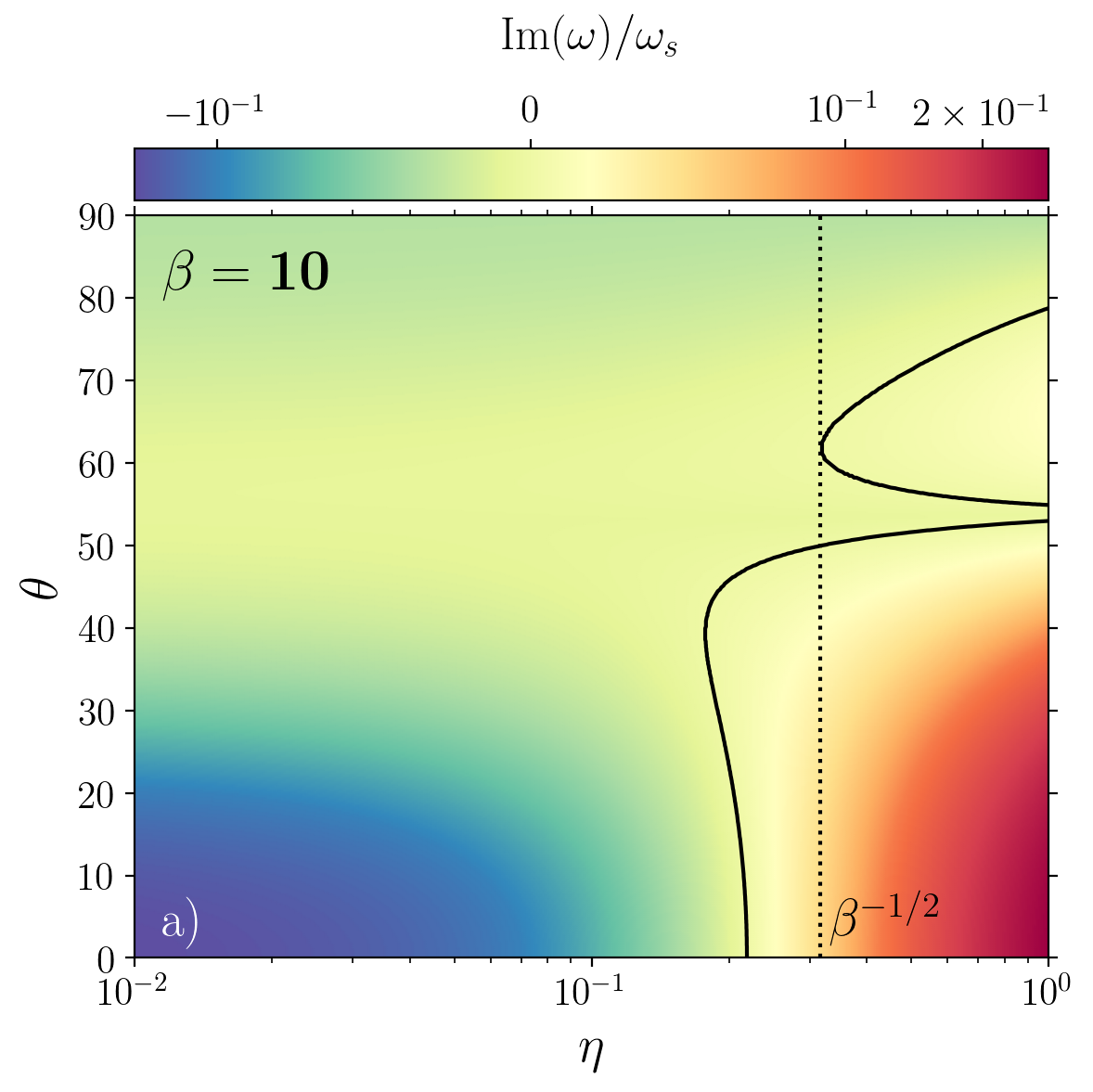}
  \end{minipage}
      \begin{minipage}[b]{0.32\textwidth}
    \includegraphics[width=\textwidth]{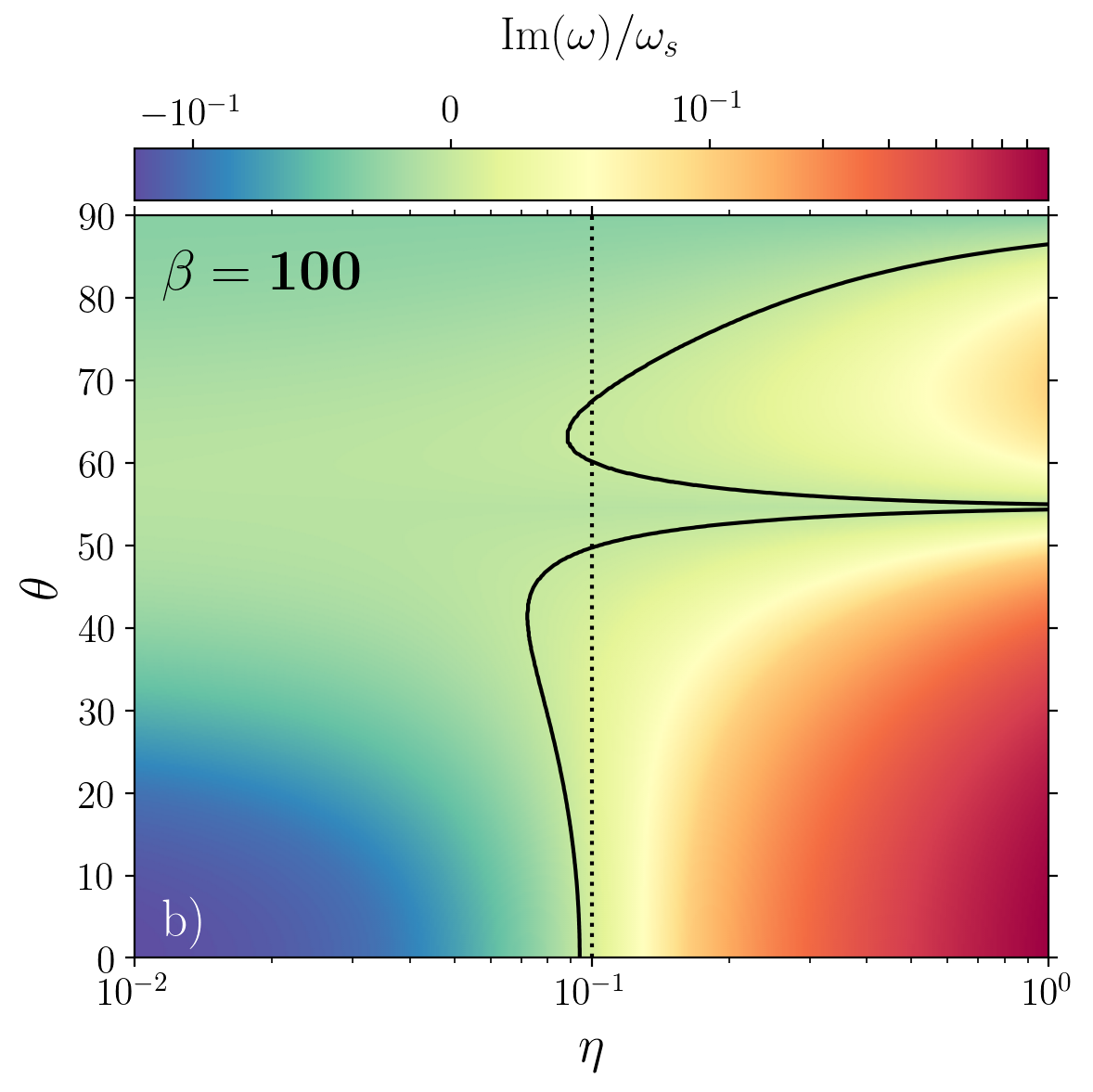}
  \end{minipage}
    \begin{minipage}[b]{0.32\textwidth}
    \includegraphics[width=\textwidth]{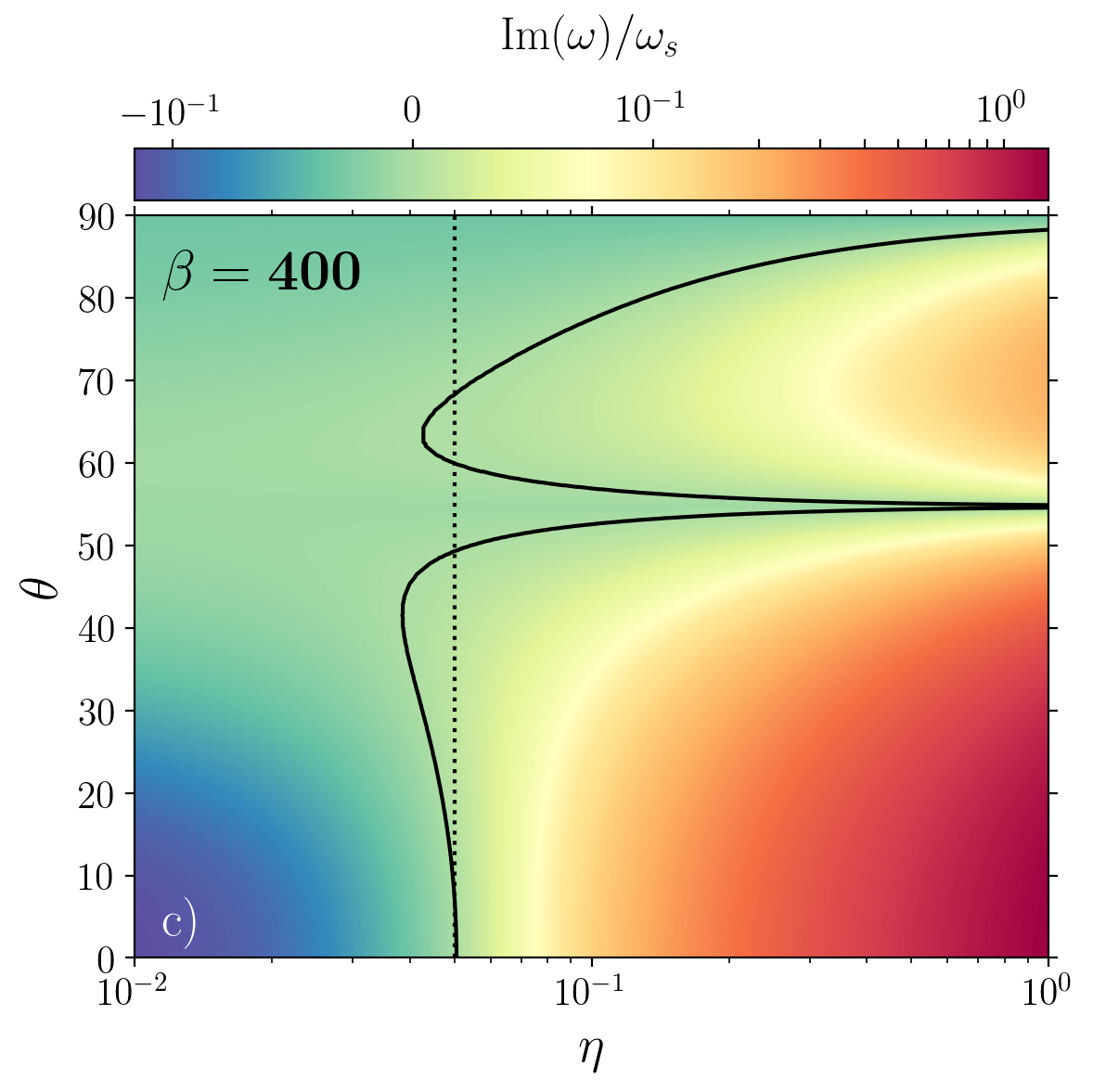}
  \end{minipage}
  \caption{Growth rates of the CR-driven acoustic instability in the $(\eta = p_c / p_g, \theta)$ plane for $\beta = 8 \pi p_g / B^2 = 10, 100, 400$, ${\rm Pr } = 1$, $\Phi = 0$ and $\nub k^2 = 0.2 \omega_s$ ($k l_{\nub} = 0.2 \sim k l_{\rm mfp}$, where $l_{\rm mfp}$ is the ion mean free path; see eq. \ref{eq:lnub}). $\imw>0$ corresponds to wave growth and the contour lines show the boundary between damping by anisotropic viscosity and the growth driven by the CRs. Marginal stability occurs first for oblique modes, but otherwise parallel propagating modes are fastest-growing. The dotted vertical lines show $\eta = \beta^{-1/2}$, which is the approximate instability-threshold scaling at high $\beta$ (e.g., eq. \ref{eq:strong_cond_eta_cond}). Even for $\eta$ well above the instability threshold (shown by the contour line), there is a ridge of stability around $\theta = 55^\circ$. For $\theta \lesssim 55^\circ$, the mode with ${\rm Re}(\omega) \approx \omega_s$ is unstable, while for $\theta \gtrsim 55^\circ$, the counterpropagating mode with ${\rm Re}(\omega) \approx - \omega_s$ is unstable (see Section \ref{sec:driving}). All colormaps in this work have log-linear  scales that are linear between $-0.1$ and $0.1$. \label{fig:fast_theta_map}}
\end{figure*}

The transition at $\theta \approx 55^\circ$ is clearly present in Figure \ref{fig:fast_theta_map}, where we show growth rates of the sound-wave instability in the $(\eta, \theta)$ plane, for $\nub k^2 = 0.2 \omega_s$ ($k l_{\nub} = 0.2$) and  $\beta  = 10, 100, 400$. Even for $\eta$ well above the instability threshold (shown by the contour line), there is a ridge of stability around $\theta = 55^\circ$. For $\theta \lesssim 55^\circ$, the mode with ${\rm Re}(\omega) \approx \omega_s$ is unstable, while for $\theta \gtrsim 55^\circ$, the counterpropagating mode with ${\rm Re}(\omega) \approx - \omega_s$ is unstable.

\subsection{1D Dispersion Relation} \label{sec:1D}

 \begin{figure*}
  \centering
      \begin{minipage}[b]{0.325\textwidth}
    \includegraphics[width=\textwidth]{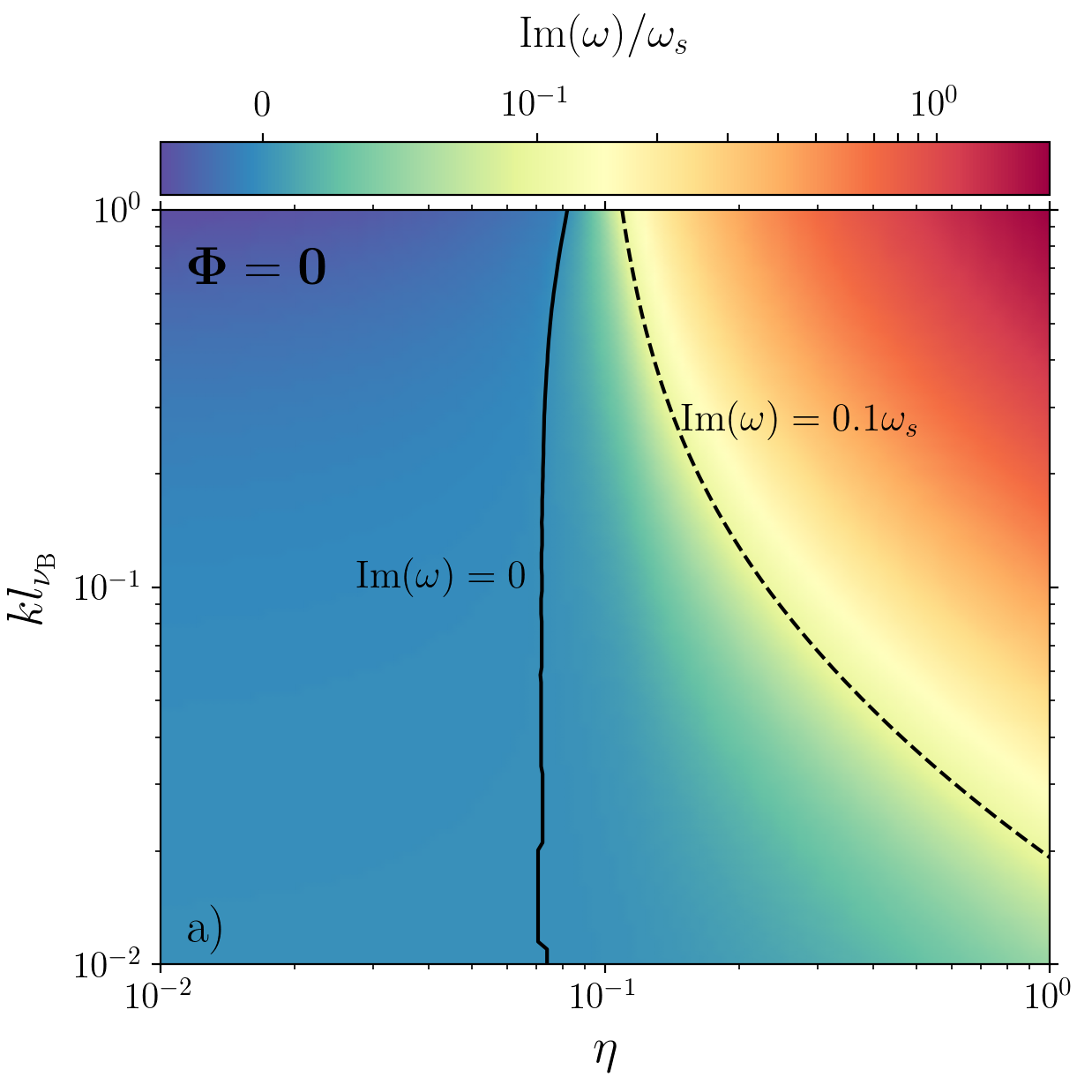}
  \end{minipage}
      \begin{minipage}[b]{0.325\textwidth}
    \includegraphics[width=\textwidth]{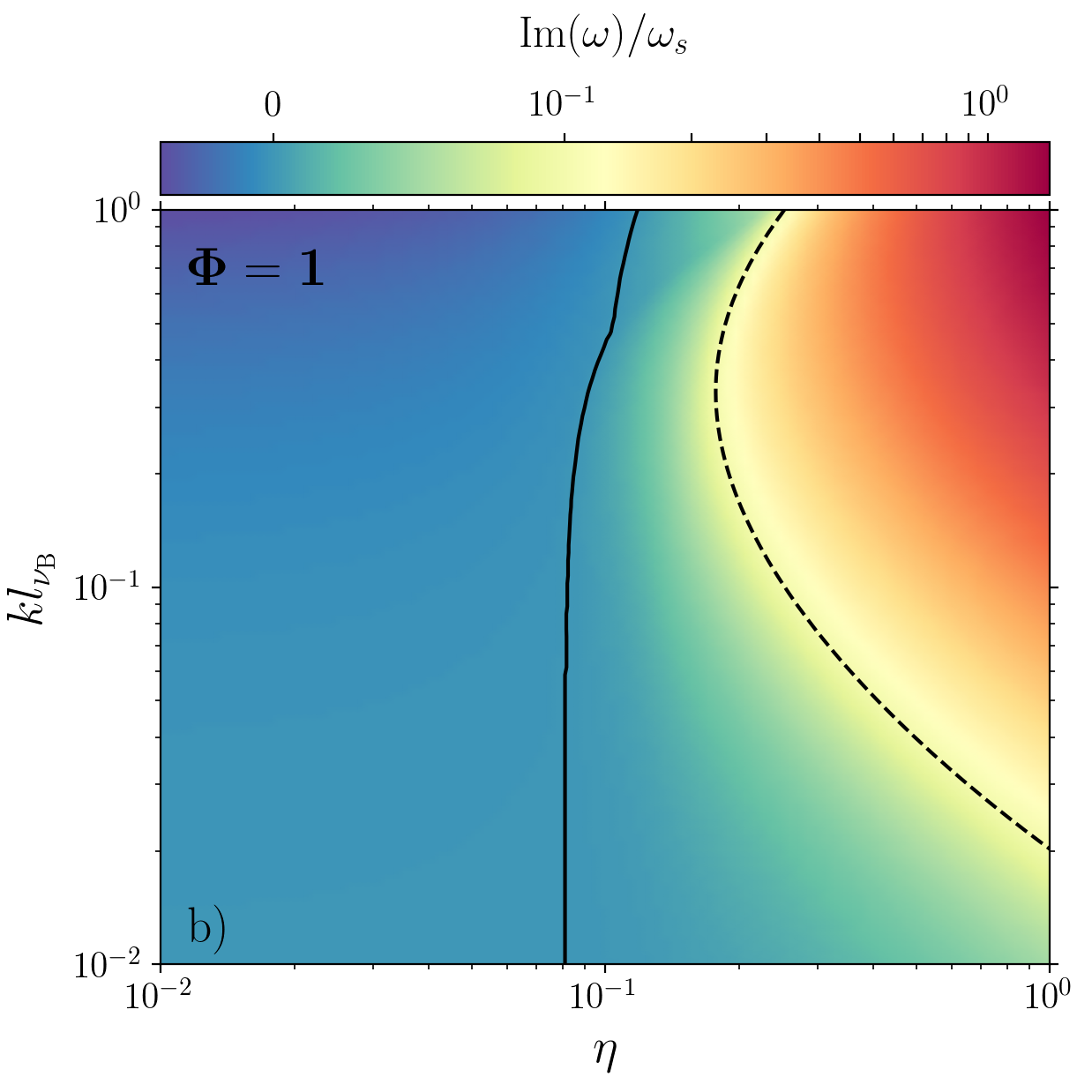}
  \end{minipage}
      \begin{minipage}[b]{0.325\textwidth}
    \includegraphics[width=\textwidth]{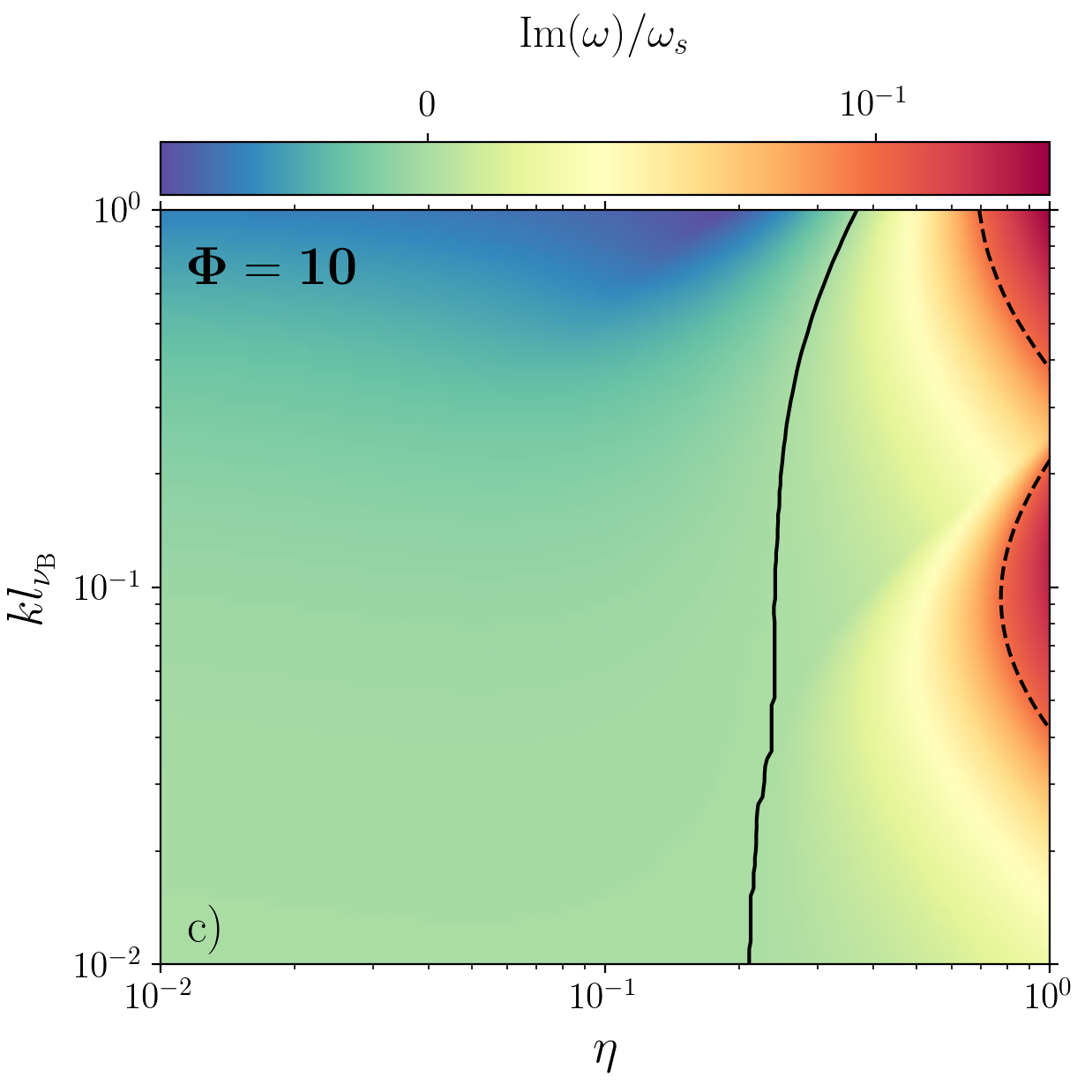}
  \end{minipage}
  \caption{ Wavenumber dependence of growth rates of the acoustic instability as a function of $\eta = p_c / p_g$ ($\beta = 100$ and $\rm Pr = 1$ in all panels). On the y-axis, $l_{\nub} \equiv \nub / c_s \sim l_{\rm mfp}$, where $l_{\rm mfp}$ is the ion mean free path (i.e. the y-axis can be written as $kl_{\nub} = \nub k^2 / \omega_s$). We consider different CR diffusion coefficients (see eq. \ref{eq:phi} for the definition of $\Phi$). The solid line corresponds to ${\rm Im}(\omega) = 0$, the dashed line is ${\rm Im}(\omega) = 0.1 \omega_s$. At each $k$, we plot the fastest growth rate (i.e. fastest growing mode across all directions of propagation, $\theta$). See Sections \ref{sec:isothermal}, \ref{sec:adiabatic} and \ref{sec:fast_crdiff} for more discussion of how $\Phi$ affects the growth rates.  \label{fig:fast_k_map}}
\end{figure*}

Because sound waves are primarily longitudinal, it is instructive and also physically well motivated to look at the instability in the 1-dimensional case. This also turns out to be sufficient to predict the approximate growth rate of the fastest growing mode in most cases, as fastest growth typically occurs for propagation parallel to $\bm{B}$. As we will show, this is not true for $\eta$ just above marginal stability, where fastest growth can occur at finite $\theta$, and when the CR diffusion coefficient is large.

For simplicity, we consider sound waves in the high-$\beta$ limit, such that $\omega \sim \omega_s \gg \omega_a$. Equation \ref{eq:dpc} then simplifies to
\begin{equation} \label{eq:phase_shift}
    \frac{\delta p_c }{p_c} = \frac{\delta \rho}{\rho} \Big(\frac{4}{3} + \frac{8}{3}i \frac{\omega_{\rm B}}{\omega_a} \Big) \Big( 1 + i \frac{\omega_d}{\omega} \Big)^{-1}.
\end{equation}
The phase shift between the CR pressure and gas density introduced by $\Delta p$ (the first bracket multiplying $\delta \rho / \rho$) is what destabilises the wave.  In contrast, the phase shift introduced by CR diffusion (second term in the second bracket) acts as a damping. 

In the high-$\beta$ limit ($\omega_s \gg \omega_a$), the 1D dispersion relation for sound waves is given by,
\begin{gather}
\begin{aligned} \label{eq:1d_complete}
    0 = & \ \omega^2 - \frac{\omega_s^2}{\gamma} \frac{\gamma \omega + i (\gamma -1) \omega_{\rm cond}}{\omega + i (\gamma-1) \omega_{\rm cond}} + \frac{4}{3} i \omega_{\rm B} \omega   \\ & - \eta \frac{\omega_s^2}{\gamma} \Big(\frac{4}{3} + \frac{8}{3}i \frac{\omega_{\rm B}}{\omega_a} \Big)\Big( 1 + i \frac{\omega_d}{\omega} \Big)^{-1} .
\end{aligned}
\end{gather}
The second term represents the standard sound-wave frequency in the presence of anisotropic conduction (adiabatic without conduction, isothermal in the limit of rapid conduction), the third term is the damping by anisotropic viscosity and the fourth term is the additional pressure response that comes from the cosmic rays, which can be destabilising. 

We first consider equation \ref{eq:1d_complete} without CR diffusion, i.e. $\omega_d = 0$. We then look at the impact of CR diffusion in Section \ref{sec:fast_crdiff}. 

\subsubsection{Nearly Isothermal Sound Waves} \label{sec:isothermal}
In the limit of rapid conduction, $\omega_{\rm cond} \gg \omega_s$ (heat conduction carried by electrons and equilibrated with the ions, i.e. ${\rm Pr} \ll 1$), the dispersion relation is (in the absence of CR diffusion)
\begin{equation} \label{eq:1d_rapid_cond}
    \omega^2 - \frac{\omega_s^2}{\gamma}  + \frac{4}{3} i \omega_{\rm B} \omega - \eta \frac{\omega_s^2}{\gamma} \Big(\frac{4}{3} + \frac{8}{3}i \frac{\omega_{\rm B}}{\omega_a} \Big) = 0.
\end{equation}
Driving by  $\delta p_c$ (Section \ref{sec:driving}) wins over damping by anisotropic viscosity when
\begin{equation}
    \eta \frac{8\omega_s^2 \omega_{\rm B}}{3 \gamma \omega_a} \gtrsim \frac{4}{3} \omega_{\rm B} \omega \approx \frac{4 \omega_{\rm B} \omega_s}{3 \sqrt{\gamma}},  
\end{equation}
where we ignored $\mathcal{O}(\eta)$ corrections to the sound speed due to the cosmic rays. The condition for instability can be written in terms of $\eta$ and $\beta$ as (in 1D):
\begin{equation} \label{eq:strong_cond_eta_cond}
    \eta \gtrsim 0.7 \beta^{-1/2} \qquad \qquad \qquad \quad {\rm (nearly \ isothermal)}.
\end{equation}
 Note that the instability threshold is independent of $\omega_{\rm B}$, as $\omega_{\rm B}$ is the characteristic frequency of both anisotropic viscous damping and the driving by $\delta p_c$. We will show that the instability threshold is generally at slightly lower $\eta$ if oblique propagation is included.

\subsubsection{ Nearly Adiabatic Sound Waves} \label{sec:adiabatic}


 If the thermal Prandtl number is not set by electron conduction and we instead have ${\rm Pr} \sim 1$, the appropriate limit to consider is $\omega_s \gg \omega_{\rm B} \sim \omega_{\rm cond}$. The dispersion relation is then approximately given by
 \begin{equation} \label{eq:disp_pr1}
     \omega^3 - \omega_s^2 \omega + i \omega_s^2 \frac{(\gamma - 1)^2}{\gamma}  \omega_{\rm cond} + \frac{4}{3} i \omega_{\rm B} \omega^2 - \eta \frac{\omega_s^2 \omega}{\gamma} \Big(\frac{4}{3} + \frac{8}{3}i \frac{\omega_{\rm B}}{\omega_a} \Big) \approx 0.
 \end{equation}
Now the driving from the $\delta p_c$ response has to compete against damping by both anisotropic conduction and viscosity (third and fourth terms, respectively). For $\omega_{\rm cond} \sim \omega_{\rm B}$ (${\rm Pr} \sim 1$), however, the correction to the instability threshold is at most order unity, and $\eta \gtrsim \beta^{-1/2}$ (eq. \ref{eq:strong_cond_eta_cond}) is still the approximate instability condition.
 
 Figure \ref{fig:fast_theta_map} shows that this $\eta \gtrsim \beta^{-1/2}$ scaling for instability works well for a wide range  of $\beta$ ($\eta = \beta^{-1/2}$ is shown by the dotted vertical lines). The plots are for $\nub k^2 = 0.2 \omega_s$ and ${\rm Pr = 1}$, i.e. $ \omega_s \gg \omega_{\rm B}, \omega_{\rm cond}$. The contours show the transition from damping by Braginskii viscosity to growth driven by the cosmic rays. Marginal stability occurs first for oblique modes, but otherwise parallel propagating modes are fastest-growing.

Figure \ref{fig:fast_k_map}a shows how growth rates depend on wavenumber $k$, for ${\rm Pr = 1}$, $\beta = 100$ and $\Phi = 0$ (no CR diffusion). The solid line corresponds to ${\rm Im}(\omega) = 0$, the dashed line is $\imw = 0.1 \omega_s$. At each $k$, we plot the fastest growth rate (i.e. fastest growing mode across all directions of propagation, $\theta$). The instability threshold is nearly independent of $k$, as the damping rates by conduction and viscosity are comparable at all $k$, so that $\eta \gtrsim \beta^{-1/2}$ is sufficient for instability across the entire range in $k$ (equation \ref{eq:strong_cond_eta_cond} and discussion in the paragraph following eq. \ref{eq:disp_pr1}). However, the growth rates generally increase with increasing $k$.

\subsubsection{Effect of CR Diffusion} \label{sec:fast_crdiff}
In the limit where CR diffusion is slow compared to the sound frequency, $\omega_d \ll \omega_s$ (this corresponds to $\Phi \ll (kl_{\rm mfp})^{-1}$), the CR term driving the instability in equation \ref{eq:1d_complete} is mildly  reduced (compared to the $\omega_d = 0$ case):
\begin{equation}
    \eta \frac{\omega_s^2}{\gamma} \frac{8}{3}i \frac{\omega_{\rm B}}{\omega_a} \rightarrow  \eta \frac{\omega_s^2}{\gamma} \Big( \frac{8}{3}i \frac{\omega_{\rm B}}{\omega_a} -\frac{4}{3}i \frac{\omega_d}{\omega} \Big).
\end{equation}
CR diffusion acts to oppose the $\Delta p$ perturbations in the CR pressure equation that drive the instability,  and as a result  shifts the instability threshold  to larger $\eta$ (compared to, e.g., eq. \ref{eq:strong_cond_eta_cond}). This shift is small, however, if $\omega_d / \omega_s \ll 2 \omega_B / \omega_a$, i.e. $\Phi \ll 2 \sqrt{\beta}$ (as well as $\Phi \ll (kl_{\rm mfp})^{-1}$, i.e. the weak diffusion limit).

In the limit $\omega_d \gg  \omega \sim \omega_s$, the CR term in eq. \ref{eq:1d_complete} is:
\begin{equation}
    \eta \frac{\omega_s^2}{\gamma} \Big( \frac{4}{3}+ \frac{8}{3}i \frac{\omega_{\rm B}}{\omega_a} \Big) \Big( 1 + i \frac{\omega_d}{\omega} \Big)^{-1} \approx  \eta \Big( -\frac{4}{3}i \frac{\omega}{\omega_d}+ \frac{8}{3}\frac{\omega_{\rm B}}{\omega_a} \frac{\omega}{\omega_d} \Big).
\end{equation}
The driving by CR pressure is completely shut off as $\delta p_c$ is suppressed by diffusion. In the 1D case considered here, for $\omega_d \gg \omega_s$ instability can only occur if $\omega \gg \omega_s$, i.e. the sound speed is much larger than  the thermal adiabatic sound speed. This occurs at $\eta \gg 1$, when the CRs set the sound speed (the CR sound speed is $\sqrt{4p_c / 3 \rho} \ $). 

Note, however, that even if $\omega_d \gg \omega_s$ for parallel propagation,  $\omega_d$ will be less than $\omega_s$ at the same $k$ for $\theta$ close to 90 degrees. As a result, for $\eta \lesssim 1$ and large CR diffusion coefficients, $\Phi \gg 1$,  the short-wavelength perturbations with $\kappa k^2 \gg k c_s$ can still be unstable for oblique propagation (this can, e.g., be seen in Figure \ref{fig:fast_fastest_eta}c).

The effects of CR diffusion as a function of wavenumber $k$ and CR pressure fraction $\eta$ are shown in Figure \ref{fig:fast_k_map}b and Figure \ref{fig:fast_k_map}c. As before, the solid line corresponds to $\imw = 0$, the dashed line is $\imw = 0.1 \omega_s$. At each $k$, we plot the fastest growth rate across all propagation angles. All parameters are the same as in Figure \ref{fig:fast_k_map}a, except for $\Phi$, which now is $\Phi = 1$ in \ref{fig:fast_k_map}b and $\Phi = 10$ in \ref{fig:fast_k_map}c. The $\Phi = 1$ growth rates are quite similar to $\Phi = 0$ (no diffusion). Noticeable differences occur primarily at high $k$, so that the overall instability threshold is not significantly changed. When CR diffusion is strong ($\Phi = 10$), significantly larger $\eta$ is required for instability. Note that there are then two regions of instability. The high-$k$ region occurs at oblique propagation, while the low-$k$ region occurs at parallel propagation.


 \subsection{Stability versus Instability \& Maximum Growth Rate}
 \label{sec:stabl_vs_instabl}
  \begin{figure}
  \centering
    \begin{minipage}[b]{0.41\textwidth}
    \includegraphics[width=\textwidth]{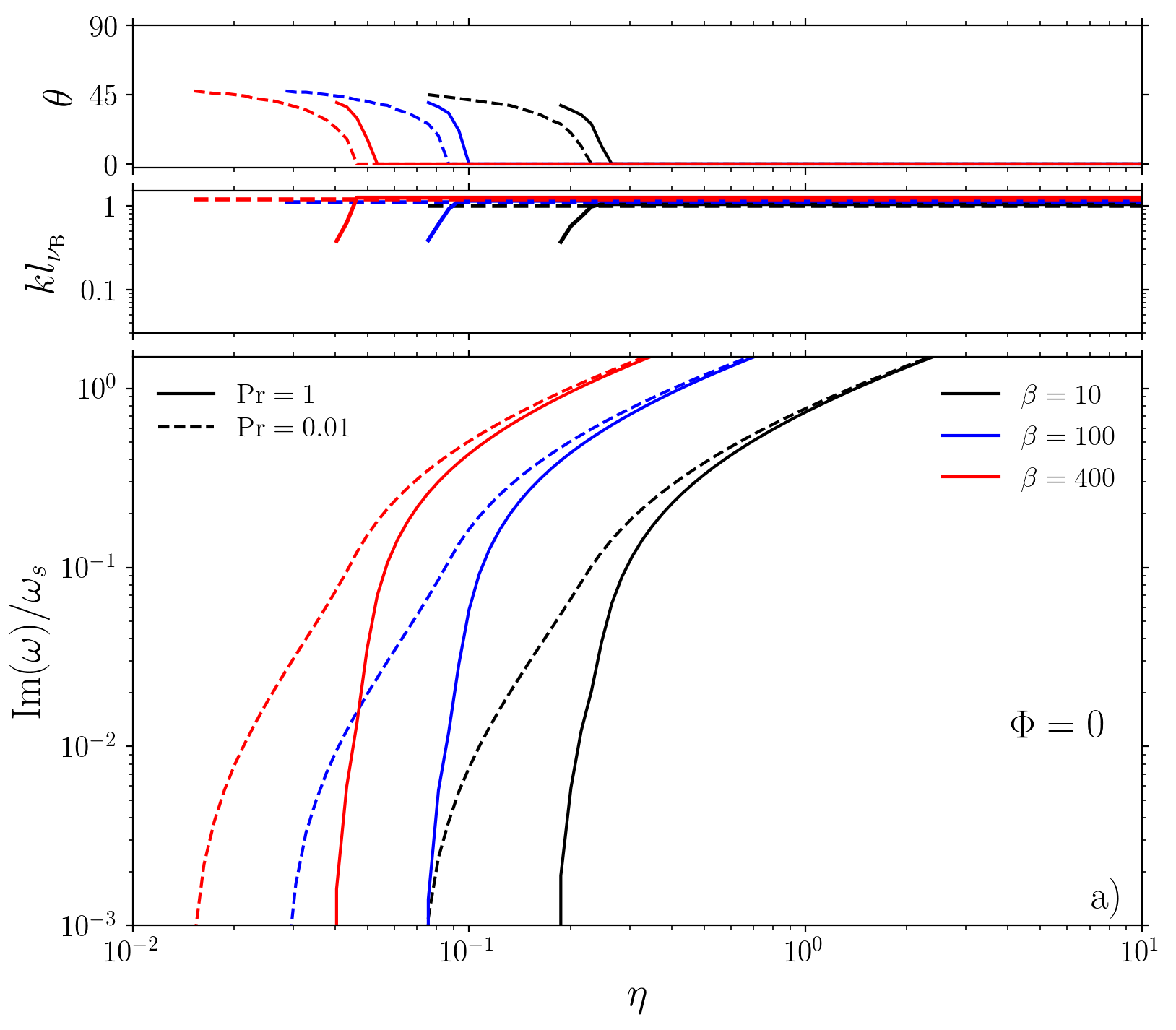}
  \end{minipage}
      \begin{minipage}[b]{0.41\textwidth}
    \includegraphics[width=\textwidth]{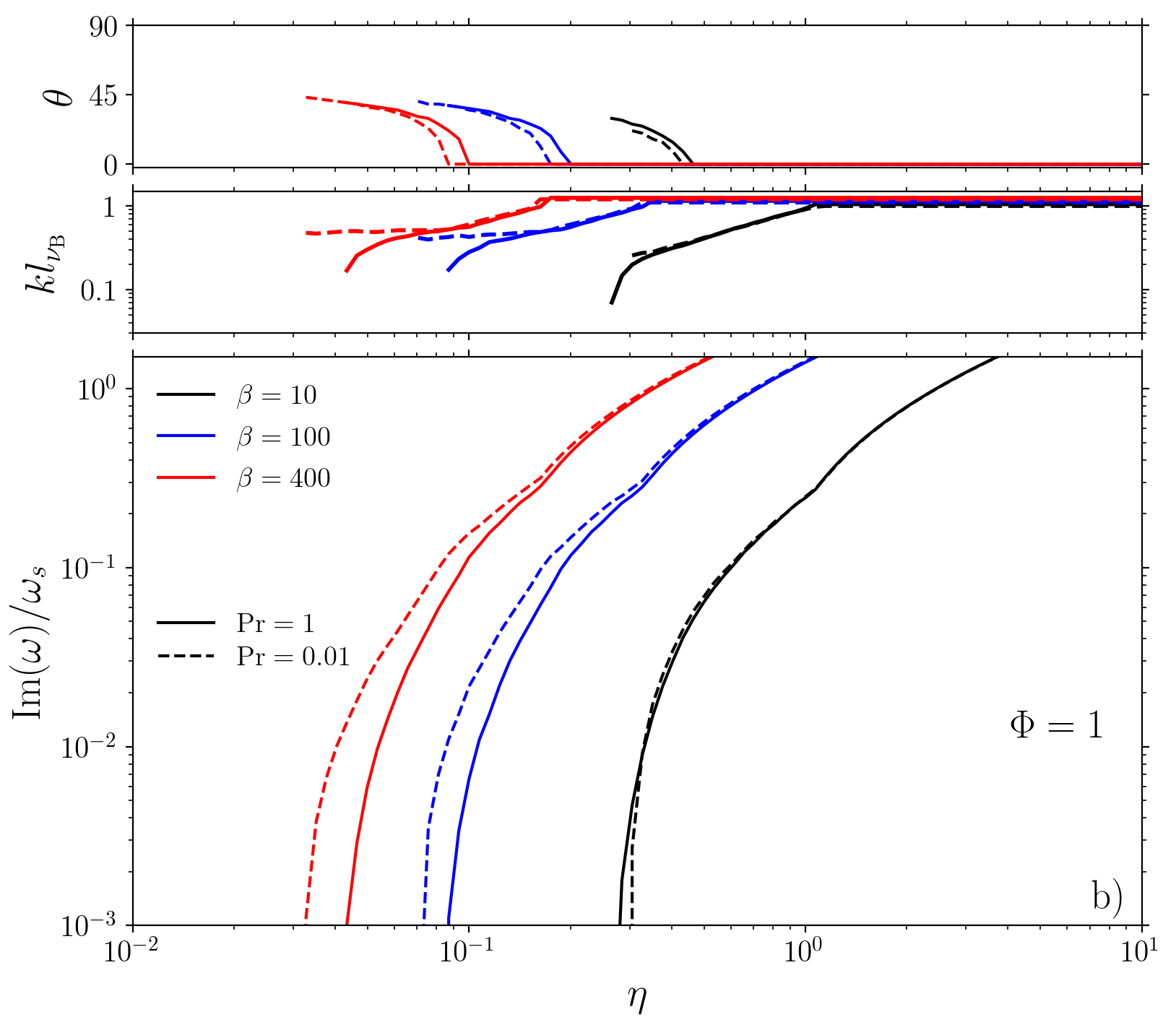}
  \end{minipage}
        \begin{minipage}[b]{0.41\textwidth}
    \includegraphics[width=\textwidth]{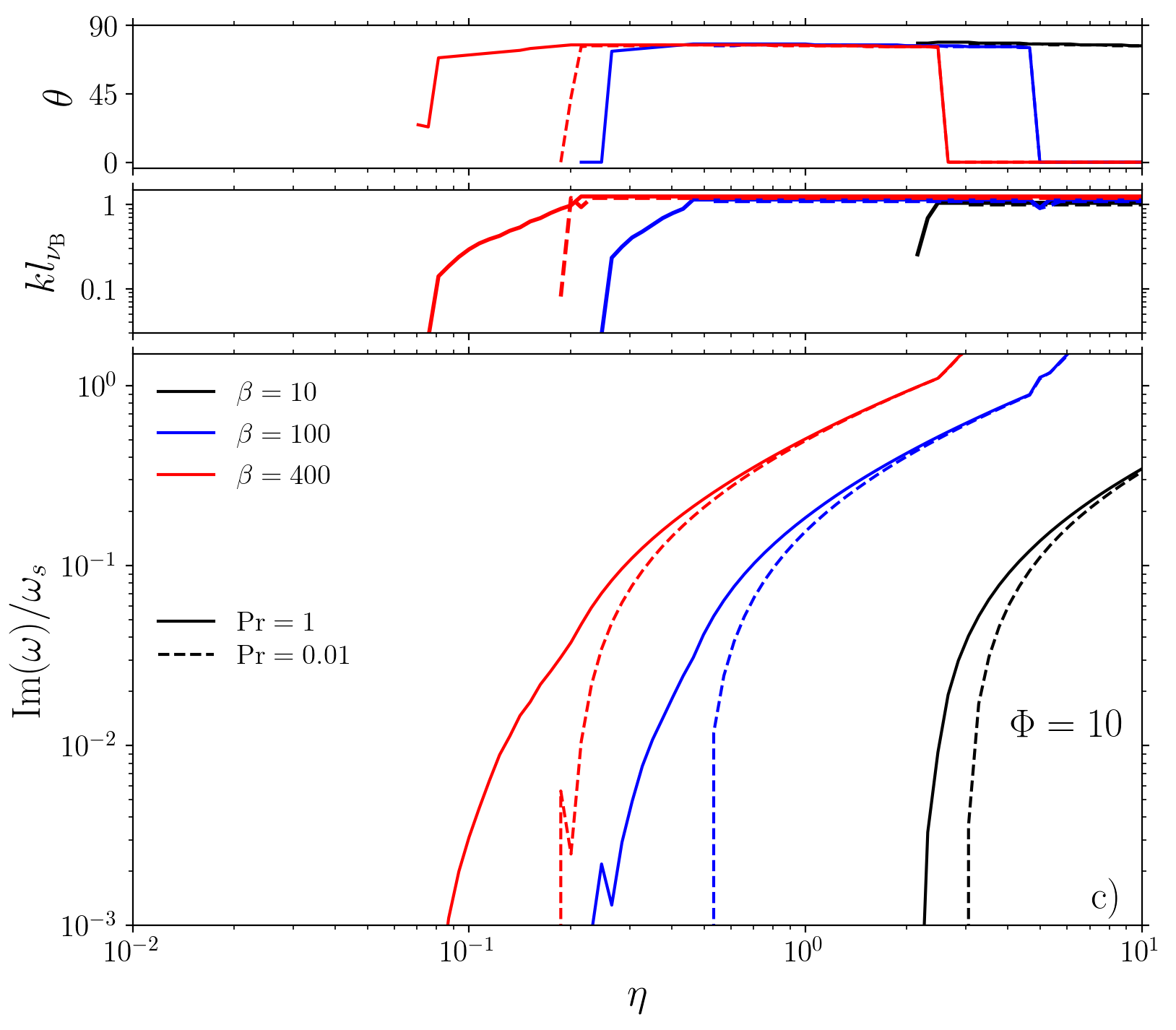}
  \end{minipage}
  \caption{Fastest growing mode of the acoustic instability as a function of $\eta$. We consider wavelengths that satisfy $k l_{\nub} \leq 1$, i.e. $\nub k^2 \leq \omega_s$ ($k  l_{\rm mfp} \lesssim 1$). We select the mode with the maximum $\imw$, but in the plots we normalise its growth rate using the adiabatic sound frequency at the $k$ where the maximum growth occurs. Panel a)  is for $\Phi = 0$ (no CR diffusion), panel b) is for $\Phi = 1$ ($\omega_d = \omega_{\rm B}$) and Panel c) is for $\Phi = 10$ ($\omega_d = 10 \omega_{\rm B}$). In each of the three panels, we also show the wavenumber $k$ and direction of propagation $\theta$ of the fastest growing mode (at $k l_{\nub} = 1$ the lines are slightly displaced for visualisation purposes). Instability occurs for smaller $\eta$ when the thermal Prandtl number ${\rm Pr}$ is small and when there is no CR diffusion. When CR diffusion is strong ($\Phi=10$)  significantly larger $\eta$ are required for instability. See Section \ref{sec:stabl_vs_instabl} for more discussion. \label{fig:fast_fastest_eta}}
\end{figure}
 
We show the fastest growing mode as a function of $\eta$, restricting to modes with $k l_{\nub} \leq 1$, in Figure \ref{fig:fast_fastest_eta}.  We select the mode with the maximum $\imw$, but in the plots we normalise its growth rate using the $\omega_s$ at the $k$ at which the maximum growth occurs. At each $\eta$, we also show the $\theta$ and $k$ at which the fastest growth occurs. The top panel is for $\Phi = 0$ (no CR diffusion), the middle panel is for $\Phi = 1$ ($\omega_d = \omega_{\rm B}$) and the bottom panel is for $\Phi = 10$ ($\omega_d = 10 \omega_{\rm B}$). We see that the minimum CR pressure fraction ($\eta$) required for instability is lowest for small thermal Prandtl numbers and no CR diffusion (see also Figure \ref{fig:fast_k_map}). While the instability threshold is not significantly modified when $\Phi = 1$, it occurs at significantly larger $\eta$ in the limit of strong CR diffusion, $\Phi = 10$. 

In the absence of CR diffusion (Figure  \ref{fig:fast_fastest_eta}a), fastest growth occurs at the highest $k$ and $\theta = 0$, except when $\eta$ is just above marginal stability. CR diffusion often shifts the fastest growing mode to lower $k$ (middle and bottom panels). However, even in the presence of CR diffusion, when $\eta$ is sufficiently above threshold, fastest growth again occurs at the highest $k$ and $\theta =0$. In the bottom panel ($\Phi = 10$), the apparent jumps in $\imw / \omega_s$, $k$ and $\theta$ are related to the existence of the two distinct regions of growth shown in Figure \ref{fig:fast_k_map}c: the low-$k$ region corresponds to $\theta = 0$, while the high-$k$ region has the large $\theta$ (the angle of this fastest-growing, $\theta \neq 0$, mode depends primarily on $\Phi$, and for $\Phi=10$ is $\approx 75^{\circ}$, as can be seen in \ref{fig:fast_fastest_eta}c).  

We conclude by stressing that for the wide range of parameters (${\rm Pr}, \ \Phi, \ \beta$) considered here, the instability and fast growth rates $\sim \mathcal{O}(\omega_s)$  occur even for small $\eta$ in high-$\beta$ evironments like the ICM. We also note that while we have focused on the simple case of a background equilibrium with $\Delta p = 0$, the acoustic instability will not be significantly affected by a finite background $\Delta p$ as long as the timescale over which the background $\Delta p$ changes is slow compared to the growth rate of the instability.\footnote{When the background $\Delta p$ is spatially varying, there will be an extra timescale, $\tau$, in our problem. However, as long as ${\rm Im}(\omega) \tau \gg 1$, which is reasonable for short-wavelength sound waves, the instability will not be significantly affected by the background $\Delta p$. The effect of a spatially constant $\Delta p$ is to modify the effective magnetic-tension and CR-heating terms, i.e. terms that are $\mathcal{O}(\omega_a)$ and negligible for our acoustic instability at high $\beta$.}

\subsection{Short Wavelengths and the Collisionless Limit} \label{sec:collisionless}
For the acoustic instability considered in this work, the Braginskii MHD model of the thermal plasma is only valid for timescales longer than the ion-ion collision time, i.e. wavelengths longer than the ion mean free path. To examine the acoustic instability below the mean-free-path scale (but at scales sufficiently large for the cosmic rays to be coupled to the gas), a collisionless description of the thermal plasma is necessary.  Preliminary calculations  using the CGL and Landau-fluid closures of the kinetic MHD equations (\citealt{cgl}; \citealt{snyder97}) suggest that the instability still exists below mean-free-path scales and has growth rates that are faster than in the weakly collisional limit. The mechanism driving the instability is somewhat different from the weakly-collisional regime illustrated in Figure \ref{fig:cartoon}:   the predominant driver in the collisionless limit is that at high $\mathbf{\beta}$, the pressure anisotropy can turn cosmic rays into a fluid with $\sim$negative effective adiabatic index, thus rendering sound waves unstable.\footnote{We consider the 1D case in which $\bm{\delta B}=0$ and for simplicity ignore the effect of heat fluxes on the pressure anisotropy (i.e. CGL closure). In the collisionless limit $\Delta p$ approximately satisfies, 
\begin{equation}
  \frac{1}{p_g} \frac{d \Delta p}{dt} \sim - \frac{1}{\rho} \frac{d \rho}{dt},
\end{equation}
so that $\delta \Delta p / p_g \sim - \delta \rho / \rho$. In contrast to the weakly collisional case, the relative phase shift between $\delta \Delta p$ and $\delta \rho$ is $\pi$ instead of $\pi /2$ (Figure \ref{fig:cartoon}). 
Assuming $\omega \approx \omega_s  \approx \sqrt{\beta}\omega_a$ and $\beta \gg 1$, $\delta p_c$ and $\delta \rho$ then roughly satisfy:  
\begin{equation}
    \delta p_c /p_c \sim - \sqrt{\beta} \delta \rho / \rho .
\end{equation}
Cosmic rays thus behave like a fluid with large negative ($\sim - \sqrt{\beta}$) adiabatic index. This can destabilise the sound wave. } 

We note, however, that the collisionless description of the thermal plasma coupled to a CR-pressure equation (eq. \ref{eq:pc}) is itself valid only at sufficiently large scales. It breaks down on small scales below the CR mean free path, where the CRs are no longer coupled to the thermal plasma (i.e. the CR scattering rate is no longer the fastest timescale in the problem). We defer a more detailed treatment of the collisionless limit to future work.    

\subsection{Relation to BZ94 Acoustic Instability}
The CRAB instability is very different from the low-$\beta$ acoustic instability driven by CR heating found in BZ94. BZ94 considered high-collisionality MHD, not the Braginskii MHD limit we have focused on. Moreover, the CRAB instability is not driven by CR heating, but by the work done on the cosmic rays by the pressure anisotropy of the thermal plasma (and is more unstable at high $\beta$). 
 
 Nevertheless, at low $\beta$ ($\beta < 1$) we do also find the BZ94 acoustic instability, albeit diminished by the damping by anisotropic viscosity and conduction (most strongly at short wavelengths). In addition to the BZ94 acoustic instability, at $\beta < 1$ there are still unstable modes driven by the pressure anisotropy. The $\Delta p$-driven instabilities at low $\beta$ require a more detailed discussion of the slow-mode instability discussed in Section \ref{sec:slow}, and so we defer the $\beta<1$ regime to future work.

\subsection{Role of Plasma Microinstabilities} \label{sec:fire_mirr}
Future simulations will shed light on the long-term evolution of the CRAB instability. Nevertheless, we can already anticipate that plasma microinstabilities growing at the ion gyroscale, such as the mirror and firehose instabilities, may significantly affect the instability at large amplitudes.

Both the mirror (\citealt{b66}; \citealt{h69}) and firehose (\citealt{r56}; \citealt{c58}; \citealt{p58}) instabilities are excited when the pressure anisotropy becomes comparable to the magnetic pressure: the mirror instability is excited when $\Delta p \gtrsim B^2 / 8 \pi$, while the firehose instability is excited when $\Delta p \lesssim - B^2 / 4 \pi$. Kinetic simulations have shown that these instabilities tend to pin the pressure anisotropy near the instability thresholds via increased scattering of particles through wave-particle interactions (\citealt{kss14}). 

When the acoustic waves grow to large amplitudes and the microinstabilities become important ($\delta \Delta p \sim B^2 / 8 \pi$), $\Delta p$ will no longer be set just by the fluid flow (i.e. the sound wave). Instead, it will be set by the plasma microinstabilities, which will act to pin $\Delta p$ near marginal stability. Recall that the work done by $\Delta p$ on the cosmic rays is the driver of the acoustic instability. It thus seems plausible that the role of the gyroscale microinstabilities will be to slow down (and/or perhaps ultimately suppress) the acoustic instability. 

At what sound-wave amplitudes do the plasma microinstabilities become important? For simplicity, consider an acoustic wave with $\delta \rho / \rho \gg \delta B /B$ (as is the case for the rapidly growing mode propagating parallel to $\bm{B}$). The pressure anisotropy is given by
\begin{equation}
   \delta  \Delta p = 3 \rho \nub \frac{d}{dt} \ln \frac{B}{\rho^{2/3}} \sim \rho \nub \omega_s \frac{\delta \rho}{\rho}.
\end{equation}
$\delta \Delta p \sim B^2 / 8\pi$ when
\begin{equation}
    \frac{\delta p_g}{p_g} \sim \frac{\omega_s}{\omega_{\rm B}} \beta^{-1} \sim \frac{1}{k l_{\rm mfp}} \ \beta^{-1} .
\end{equation}
 In high-$\beta$ systems, it is therefore the long-wavelength modes that can grow to large amplitudes without exciting kinetic microinstabilities.  Short-wavelength  perturbations ($kl_{\rm mfp} \sim 1$), which tend to be the fastest growing modes, are affected by pressure-anisotropy-driven microinstabilities at smaller amplitudes than the long-wavelength modes.

\subsection{Instability of the Slow-Magnetosonic/CR-Entropy Mode} \label{sec:slow}
In addition to the acoustic instability (instability of the fast-magnetosonic wave), there is also a second unstable mode in our problem, associated with the slow-magnetosonic and CR-entropy modes. The instability is driven by a fluid resonance between the CR-entropy mode and the MHD slow-magnetosonic mode, which both share a characteristic eigenfrequency $\omega_a$ at high $\beta$. At high $\beta$ the growth rates of the slow-mode (or, CR-entropy mode) instability are significantly smaller than the growth rates of the acoustic instability that is the focus of this work. For this reason, we defer a more detailed analysis of these additional instabilities to a future paper. We do point out, however, that in the absence of CR diffusion the instability of the CR-entropy mode exists for any CR pressure, $\eta \neq 0$.

\section{Applications} \label{sec:applications}
In this section we speculate on example astrophysical applications of the CR-driven acoustic instability. We first consider the impact of the CRAB instability on sound waves propagating through galaxy clusters (Section \ref{sec:perseus}). This is motivated by large-amplitude surface-brightness fluctuations observed in the Perseus cluster (\citealt{fabian03}), often interpreted to be  long-wavelength sound waves. In \ref{sec:shock} we speculate that cosmic rays may efficiently excite sound waves in the vicinity of shocks and in the outskirts of galaxy and cluster halos close to the virial radius. We also argue that the sound waves excited by the low-energy GeV cosmic rays may be important for the scattering of higher-energy cosmic rays (Section \ref{sec:confine}).

\subsection{X-Ray Ripples in Perseus} \label{sec:perseus}
 \begin{figure}
  \centering
    \begin{minipage}[b]{0.45\textwidth}
    \includegraphics[width=\textwidth]{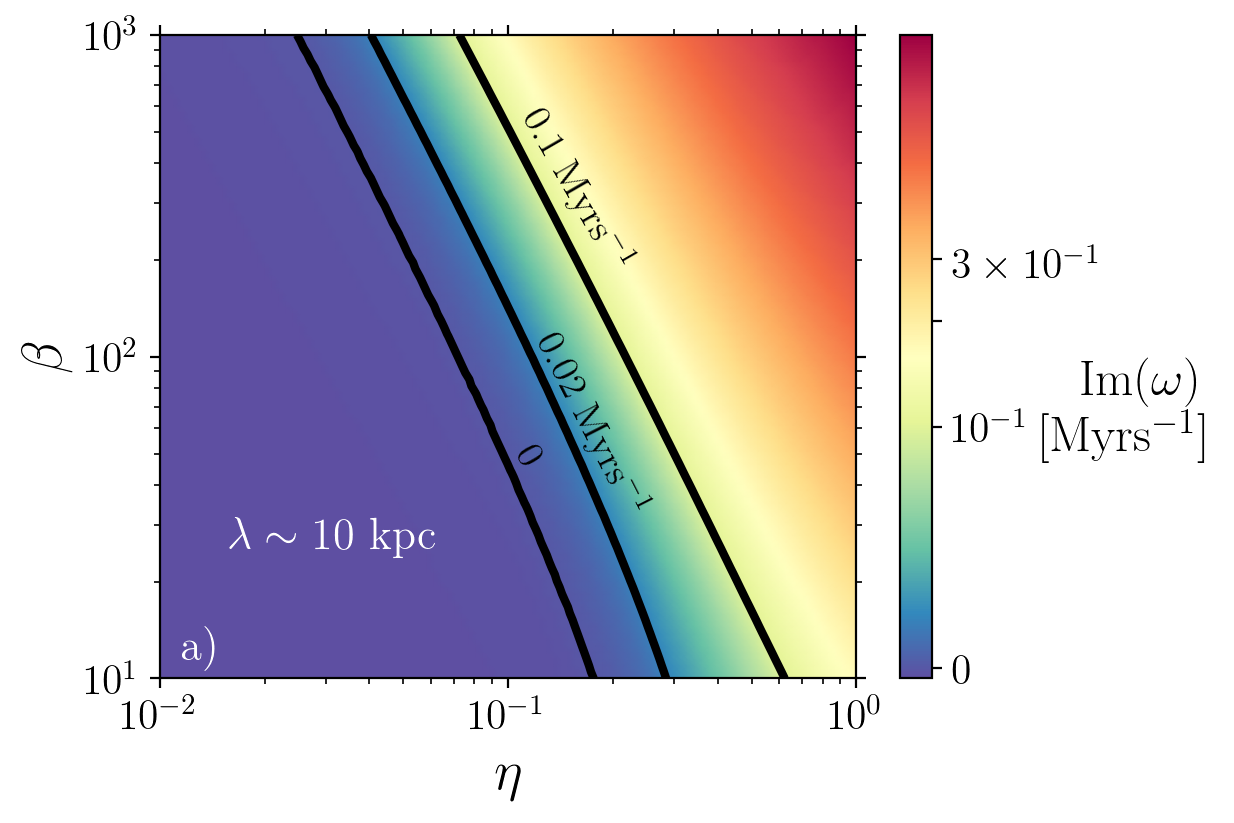}
  \end{minipage}
      \begin{minipage}[b]{0.45\textwidth}
    \includegraphics[width=\textwidth]{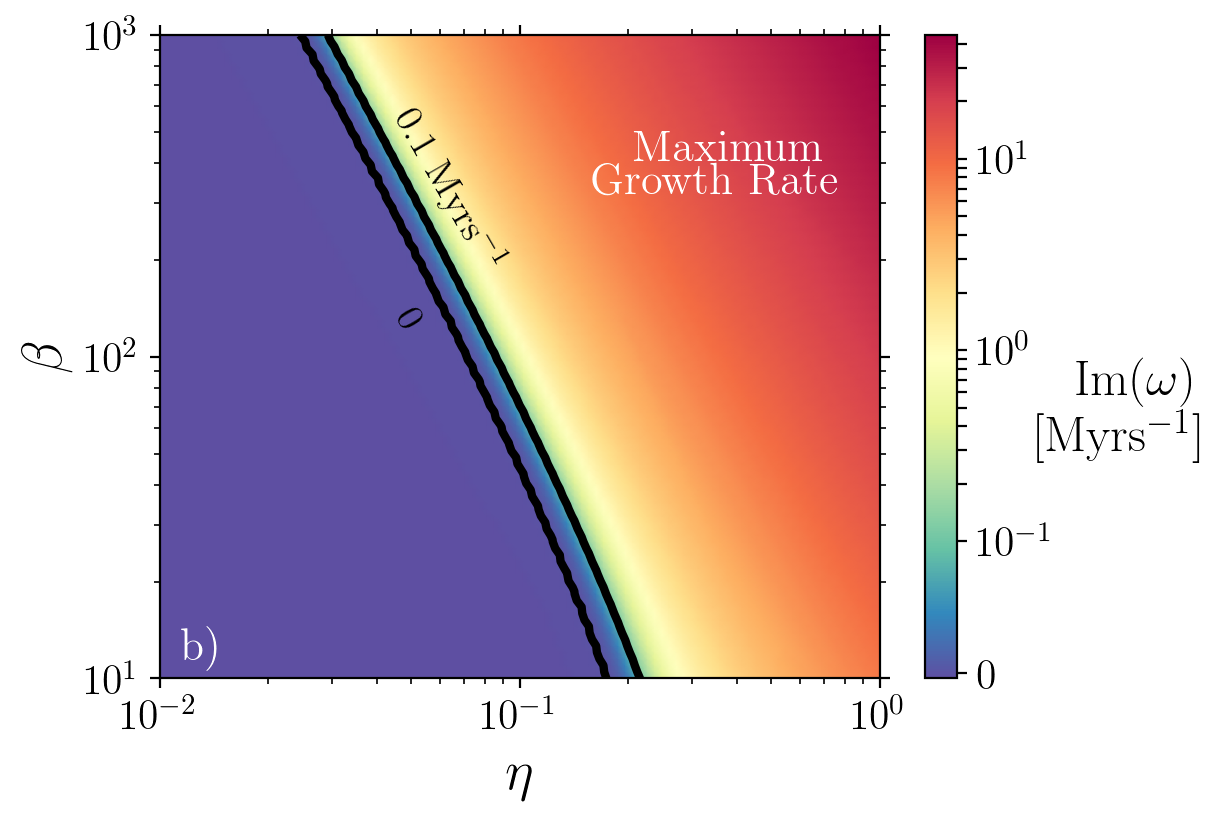}
  \end{minipage}

  \caption{Growth rate of the acoustic instability as a function of $\eta$ and $\beta$ for Perseus-like parameters: $T \sim 5 \times 10^7$K and $n_{\rm i} \sim 0.03 \  {\rm cm^{-3}}$ (\citealt{fabian06}), so that the ion mean free path $l_{\rm mfp}$ is of order  0.1 kpc. Here we assume $\beta = 100$, ${\rm Pr = 1}$ and no CR diffusion. Panel a) shows the maximum growth rate of a $k l_{\nub} = 0.1$ sound wave ($\lambda \sim 10$ kpc; 10 kpc corresponds to the approximate wavelength of the X-ray surface-brightness fluctuations in Perseus as observed by \textit{Chandra}). Note that these growth rates are larger at larger distances from the cluster center, where the density is lower and the mean free path is larger. Panel b) shows the maximum growth rate when we consider all wavelengths that satisfy $k l_{\rm mfp} \leq 1$. Significant amplification over timescales of order $10 \ {\rm Myrs}$ (timescale for sound waves to propagate tens of kpc) is plausible for realistic values of $\eta$ and $\beta$.    \label{fig:eta_beta} }
\end{figure}

\textit{Chandra} X-ray observations have revealed long-wavelength, $\mathcal{O}(10 \ {\rm kpc})$, surface-brightness ripples in the Perseus cluster (\citealt{fabian03}; \citealt{fabian06}). The inferred $\mathcal{O}(10\%)$ density fluctuations are believed to be sound waves propagating through the cluster. More generally, it is believed that sound waves excited by time-variable AGN activity are important for heating cluster plasmas (e.g., \citealt{li_bryan_15}; \citealt{bambic2019}). The gas in Perseus and other clusters is weakly collisional and is likely also filled with cosmic rays. Thus, it is plausible that these sound waves are affected by the CRAB instability described in this paper. 

The CR pressure fraction in Perseus and other cluster environments is constrained to be of order a few percent to a few tens of percent.\footnote{A few percent according to \cite{aleksic10} and \cite{aleksic12}, but their study uses primarily high-energy CRs. The upper limit on the total CR pressure in Perseus -- dominated by the low-energy CRs that are the most important for this work -- is significantly larger in \cite{huber13}.} For a gas temperature  of $5 \times 10^7$K and number density $0.03 \  {\rm cm^{-3}}$ appropriate for Perseus (\citealt{fabian06}), the ion mean free path is of order 0.1 kpc. This translates into $\omega_{\rm B} / \omega_s \sim k l_{\rm mfp} \sim 0.1$ for a $\sim$10 kpc wavelength. By how much can this wave be amplified through the CRAB instability?

We show growth rates of a $k l_{\nub} = 0.1$ (a wavelength of order $\lambda \sim 10$ kpc) acoustic wave in Myrs$^{-1}$ in the ($\eta$, $\beta$) plane in Figure \ref{fig:eta_beta}a. We use ${\rm Pr} = 1$ and assume no CR diffusion, $\Phi = 0$. To clearly show where the instability becomes important, we explicitly show contours where the growth rates are $0, 0.02$ and $0.1 \ {\rm Myrs^{-1}}$.

 The sound speed in Perseus is of order $\sim  10^8 \ {\rm cm \ s^{-1}}$, so that waves propagate a distance $50$ kpc (say) in $\sim$50 Myrs. For the wave to undergo at least one e-folding in that time, the required growth rate is $\imw \gtrsim 0.02 \ {\rm Myrs^{-1}}$. This is satisfied if $\eta$ and $\beta$ are sufficiently large, see Figure \ref{fig:eta_beta}a. Whether rapid growth of long-wavelength waves does indeed occur in Perseus is somewhat unclear, due to the lack of sufficiently good constraints on the cluster magnetic-field strength and cosmic-ray pressure (there are also uncertainties in the particle mean free path due to the role of plasma microinstabilities). However, it seems possible at least in localised regions with sufficiently large $\eta$ and $\beta$ (see also Section \ref{sec:shock}).  We also note that for $\beta \sim 100$, a $k l_{\rm mfp} = 0.1$ ($\sim 10$ kpc) wave  can grow to fairly large amplitudes, $\delta p_g / p_g \sim \delta \rho / \rho \sim 10 \%$, before pressure-anisotropy microinstabilities become important which likely slow down and/or ultimately suppress the instability (see Section \ref{sec:fire_mirr}). This is consistent with the $\mathcal{O}(10\%)$ density fluctuations inferred in Perseus (\citealt{fabian03}; \citealt{fabian06}). Finally, we note that at larger distances from the cluster core the density is lower and the mean free path is larger. As a result, long-wavelength ($\lambda \sim 10$ kpc) modes will have faster growth rates at large distances from the cluster center.  

In  Figure \ref{fig:eta_beta}b we do not restrict our attention to $\lambda \sim 10$ kpc wavelengths, and instead show the overall maximum growth rates in the $(\eta, \beta)$ plane. We use the same Perseus temperatures and densities as before, such that $l_{\rm mfp} \sim 0.1$ kpc, and we consider wavelengths satisfying $k l_{\rm mfp} \leq 1$ ($k l_{\nub} \leq 1$). The CRAB instability occurs and has fast growth rates for a wide range of realistic cluster values of $\eta$ and $\beta$. We therefore conclude that cosmic rays likely lead to large amplifications of kpc-scale sound waves propagating in dilute cluster plasmas.

\subsection{Sound-Wave Excitation in the Vicinity of (Virial) Shocks} \label{sec:shock}
The CRAB instability is particularly important at high cosmic-ray pressures, i.e. large $\eta$. This suggests that the instability is easily excited in the vicinity of shocks that are responsible for CR acceleration, i.e. where $\eta$ is typically much higher than its average value in the ambient medium. This may be relevant for shocks in supernova remnants and shocks driven by galactic winds or AGN jets in galaxy halos and clusters. 

In addition, cluster simulations that include the production of cosmic rays in structure-formation shocks find that the CR pressure fraction is higher close to the virial radius (virial shock) than in the central regions of the cluster (\citealt{pfrommer2008}). It seems possible that sound waves excited close to the virial radius through the CRAB instability discussed in this work can then propagate in towards the cluster core. Modes with longer wavelengths, $\sim$ 10s of kpc, will grow much faster at large radii near the viral radius than in the cluster core because of the much lower density and larger ion mean free path at these radii.   
\subsection{Scattering of High-Energy Cosmic Rays} \label{sec:confine}

 \begin{figure}
  \centering
    \includegraphics[width=0.45\textwidth]{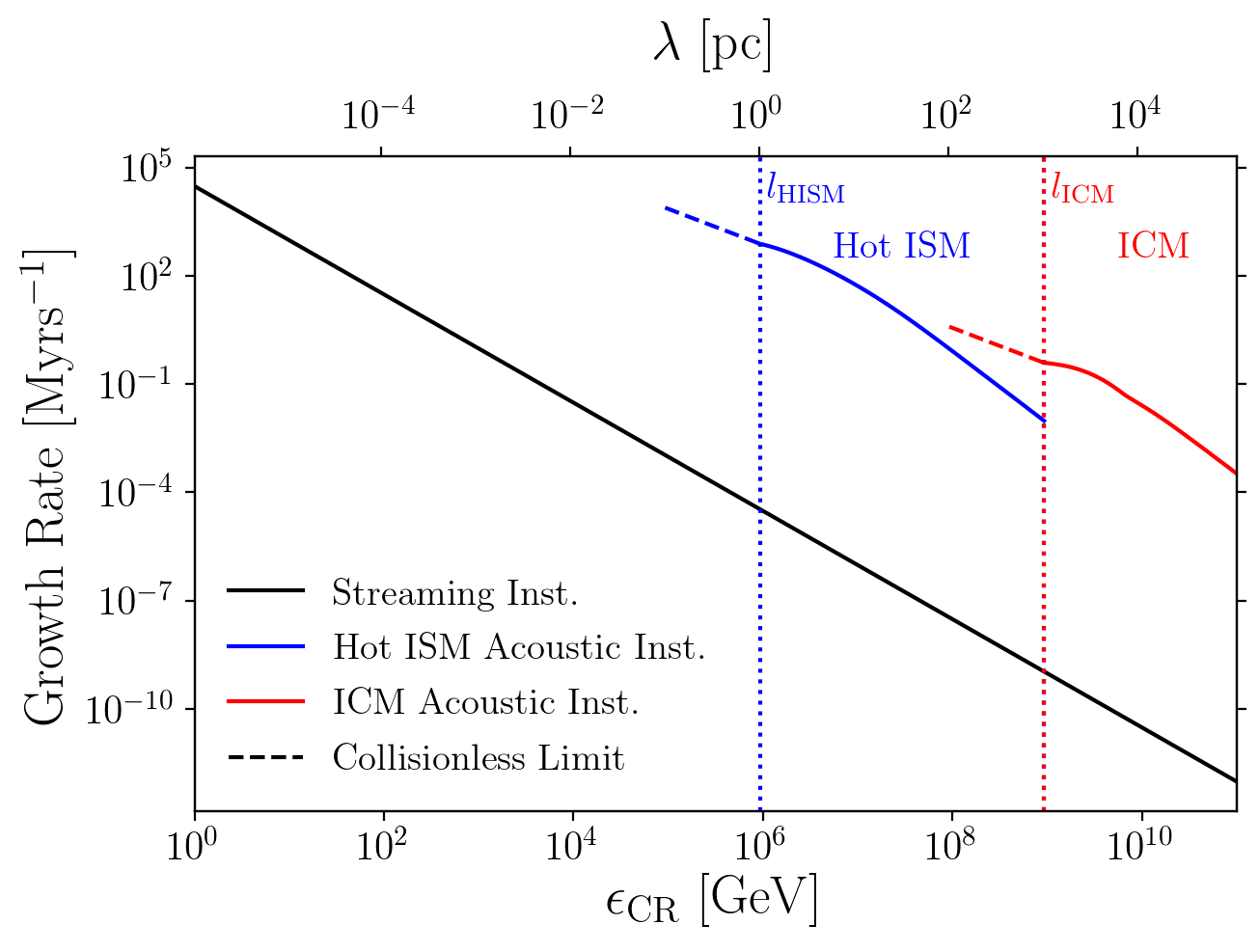}
  \caption{Schematic growth-rate comparison of the gyroresonant streaming instability of Alfv\'en waves and the long-wavelength acoustic instability excited by the GeV CR fluid coupled to the thermal plasma. The growth rate is plotted against CR energy (bottom horizontal axis) as well as wavelength (top horizontal axis; the two are related by the resonance condition  $r_{\rm L} / \lambda \sim \epsilon_{\rm CR} / \lambda e B \sim 1$, where $r_{\rm L}$ is the CR gyroradius). For the streaming-instability growth rates we use a single CR spectral slope, $\alpha = 4.5$, a 1 $\mu$G magnetic field, $n_{\rm CR} / n_{\rm i} = 10^{-7}$   and  $({\rm v_D - v_A})/{\rm v_A} = 1$ (eq. \ref{eq:gyrores}). The acoustic instability is plotted for ${\rm Pr = 1}$ and $\Phi = 0$ (no CR diffusion), and wavelengths larger than the ion mean free path, $l_{\rm mfp} / \lambda \leq 1$ (we assume $l_{\rm mfp} = 1$ pc in the hot ISM and $l_{\rm mfp} = 1$ kpc in the ICM). We use $\eta = 1, \ \beta=10$ in the hot ISM and $\eta = 0.1, \ \beta = 100$ in the ICM (these values of $\eta$ correspond to $n_{\rm CR} / n_{\rm i} \sim \mathcal{O}(10^{-7})$ for typical hot ISM and ICM temperatures). At long wavelengths, the growth rates of the acoustic instability are orders of magnitude larger than the streaming-instability growth rates. In principle, the sound-wave instability grows sufficiently fast to contribute to the scattering of higher-energy ($\sim$PeV and $\sim$EeV) cosmic rays. The dashed blue and red lines represent sub-mean-free-path scales, where the thermal plasma is collisionless. We defer a detailed treatment of this regime to future work, but preliminary calculations suggest that the instability is still present in the collisionless limit. 
   \label{fig:growths} }
\end{figure}

The overstable sound waves found in this paper have long wavelengths ($\gtrsim 1$ kpc in the ICM and $\gtrsim 1$ pc in the hot ISM) and can have growth rates that are significant compared to the oscillation frequency. The purpose of this section is to point out that the growth rates of the sound-wave instability are significantly larger than the growth rates of the Alfv\'en waves excited by high-energy CRs through the streaming instability. The streaming-instability growth rate is given by (\citealt{kp69}; \citealt{zweibel_micro}),
\begin{equation} \label{eq:gyrores}
    \Gamma_k \sim \Omega_0 \frac{n_{\rm CR} (p> p_{\rm min})}{n_{\rm i}} \frac{{\rm v_D - v_A}}{{\rm v_A}},
\end{equation}
where ${\rm v_D}$ is the CR drift speed, $\Omega_0$ is the nonrelativistic gyrofrequency and $n_{\rm i}$ is the thermal ion number density. $n_{\rm CR} (p>p_{\rm min})$ is the number density of CRs that can resonate with a wave with wavenumber $k$, and $p_{\rm min} = m \Omega_0 / k$. Because the CR spectrum is steep, the number of high-energy cosmic rays resonating with long-wavelength modes is very small. This leads to very small $\Gamma_k$ for modes with wavelengths that can scatter and confine the high-energy CRs: $f(p) \propto p^{-\alpha}$ with $\alpha \approx 4.5$, so $n_{\rm CR} (p> p_{\rm min}) \propto p_{\rm min}^{3-\alpha} \propto k^{\alpha - 3}$, which decays rapidly with CR energy. As a result, high-energy cosmic rays are not able to confine themselves. Here we inspect the possibility that the acoustic instability excited by the GeV cosmic-ray fluid can scatter and at least partially confine higher-energy cosmic rays.  

Figure \ref{fig:growths} shows a schematic growth-rate comparison of the gyroresonant streaming instability of Alfv\'en waves and the CRAB instability considered in this work. The growth rate is plotted against CR energy (bottom horizontal axis) as well as wavelength (top horizontal axis). The wavelength and CR energy are related by the resonance condition $\epsilon_{\rm CR} \sim \lambda e B$. We assume a CR spectral slope $\alpha = 4.5$ and $({\rm v_D - v_A})/{\rm v_A} = 1$. For the acoustic instability we consider wavelengths larger than the ion mean free path, $l_{\rm mfp}/ \lambda \leq 1$. For the hot ISM, we assume an ion mean free path $l_{\rm mfp} = 1$ pc and for the ICM we assume an ion mean free path $l_{\rm mfp} = 1$ kpc. The Braginskii MHD description of the thermal plasma is appropriate above the ion mean-free-path scale. However, preliminary calculations using collisionless fluid closures suggest that the instability also exists below the mean-free-path scale (see Section \ref{sec:collisionless}). We show this using the dashed blue and red lines. We stress again that the collisionless description of the thermal plasma coupled to a CR-pressure equation breaks down at small scales where the CRs are no longer coupled to the thermal plasma. The growth rates are not plotted below this scale in Figure \ref{fig:growths} (the CR mean free path is somewhat uncertain and for this reason we extend growth rates only one order of magnitude below the ion mean-free-path scale; however, this range might be significantly larger, e.g. in the ICM where the ion mean free path is large). 


 Figure \ref{fig:growths} shows that the growth rates of the CRAB instability are orders of magnitude faster than the growth rates of the streaming instability excited by the high-energy CRs. The growth rate is relatively independent of propagation angle for $\theta \lesssim 55 ^\circ$ (Figure \ref{fig:fast_theta_map}), so modes with appreciable $\delta B_\perp / B$ can be excited. Sound waves may, in principle, grow sufficiently fast to reach large amplitudes and efficiently scatter high-energy cosmic rays. The CR scattering rate is proportional to $\Omega (\delta B_\perp / B)^2$. If the acoustic waves destabilised by the GeV CRs saturate at sufficiently large $\delta B_\perp / B$, the acoustic instability identified here may significantly affect cosmic-ray confinement. While large $\delta B_\perp / B$  seem possible given the fast growth rates, future simulations will be necessary to study the saturation of the instability and address the efficiency of scattering high-energy CRs. Finally, we note that turbulence will likely be produced in the gas as a result of the CRAB instability. This may significantly affect the scattering and transport properties of intermediate-energy ($\lesssim {\rm PeV}$) cosmic rays, whose gyroradii are too small to directly resonate with linearly unstable acoustic waves. 


\section{Conclusions} \label{sec:conclusions}
The interstellar, circum-galactic and intracluster media are filled with dilute, weakly-collisional plasmas characterised by anisotropic viscosity and conduction. Without cosmic rays, these anisotropic transport properties lead to the well-known damping of sound waves (the slow and fast magnetosonic modes). In this paper we have shown that when cosmic rays are present, sound waves can instead grow exponentially in time, even for small CR pressures. We have termed this the Cosmic Ray Acoustic Braginskii (CRAB) instability.

We model the dilute plasmas filled with cosmic rays by using the Braginskii MHD  closure for weakly collisional plasmas (\citealt{br65}) coupled to a pressure equation for the cosmic rays (Section \ref{sec:equations}). The cosmic rays are assumed to stream at the Alfv\'en speed $\vadp$, which in a weakly collisional plasma depends on the pressure anisotropy $\Delta p$ (eq. \ref{eq:va_mod}). We also include CR diffusion along the magnetic-field direction.

The key frequencies and dimensionless parameters in our problem are summarised in Section \ref{sec:params}. We  focus on high-$\beta$ ($\beta= 8 \pi p_g / B^2 \sim 100$) plasmas, as is appropriate for the ICM. The Braginskii MHD model is valid provided that the timescales of interest are longer than the ion-ion collision time. We impose this by   constraining the anisotropic-viscous (Braginskii) frequency, $\omega_B$, to be smaller than the sound frequency, $ \omega_s$ (see Section \ref{sec:params}).

The CRAB instability  is driven by a phase shift between the CR-pressure and gas-density perturbations. This phase shift is introduced by the dependence of the Alfv\'en speed on $\Delta p$ (eq. \ref{eq:va_mod}). The physical mechanism driving the instability is sketched out in Figure \ref{fig:cartoon}: work done by the pressure anisotropy on the cosmic rays enhances regions of larger than average CR pressure, leading to a positive feedback loop. Sound waves are unstable if $\eta = p_c / p_g \gtrsim \alpha \beta^{-1/2}$, where $\alpha$ depends on the thermal Prandtl number and the CR diffusion coefficient. We find that $\alpha$ is typically slightly less than $1$ (unless the CR diffusion coefficient is much larger than the thermal-plasma anisotropic viscosity, in which case $\alpha >1$; see bottom panel of Figure \ref{fig:fast_fastest_eta}). Thus, even small CR pressures are sufficient for instability in high-$\beta$ plasmas such as the ICM. We find that the acoustic instability is characterised by large growth rates, comparable to the sound-wave oscillation frequency (Figure \ref{fig:fast_fastest_eta}).   

The growth rates absent CR diffusion are not a strong function of propagation angle relative to $\bm{B}$ for $\theta \lesssim 55^\circ$ (Figure \ref{fig:fast_theta_map}). However, the fastest growing mode is typically propagating parallel to the magnetic-field direction (except for  $\eta$ just above marginal stability or when CR diffusion is strong, see Figure \ref{fig:fast_theta_map} and Figure \ref{fig:fast_fastest_eta}). This result motivated a simplified 1D derivation of the dispersion relation, which we show in \eqref{eq:1d_complete}. Growth rates are typically largest at the highest $k$, except at small $\eta$ just above the instability threshold or when CR diffusion is significant (Figure \ref{fig:fast_k_map} and Figure \ref{fig:fast_fastest_eta}).

We considered astrophysical implications of the CRAB instability in Section \ref{sec:applications}. In Section \ref{sec:perseus} we argue that the instability is likely important for amplifying sound waves propagating through galaxy cluster and group environments. This includes the Perseus cluster, where long-wavelength, large-amplitude X-ray surface-brightness fluctuations observed by \textit{Chandra} are often interpreted as sound waves. We show instability growth rates as a function of $\eta$ and $\beta$ for Perseus-like parameters in Figure \ref{fig:eta_beta}. In Section \ref{sec:shock} we hypothesise that the acoustic instability is likely important near shocks, where the CR pressure is large. This includes the outskirts of galactic and cluster halos, i.e. regions close to the virial shock, as well as shocks associated with supernovae, galactic winds, or AGN winds/jets propagating into the hot ISM or halo environments. In Section \ref{sec:confine} we speculate that the long-wavelength acoustic modes excited by the GeV cosmic-ray fluid can contribute to the scattering of higher-energy cosmic rays. In Figure \ref{fig:growths} we show that the long-wavelength acoustic modes grow orders of magnitude faster than the Alfv\'en waves excited by the high-energy CRs through the gyroresonant streaming instability. It remains to be seen, however, whether the sound waves grow to large enough amplitudes and/or generate smaller-scale fluctuations through turbulence to efficiently scatter TeV to EeV cosmic rays. 

Future simulations will address the saturation of the CRAB instability. They will show whether the excited sound waves can grow to large enough amplitudes to efficiently scatter high-energy cosmic rays. Global simulations that include both Braginskii MHD and cosmic rays will shed light on the importance of the acoustic instability for the evolution of gas and the propagation of sound waves in the ISM, galactic halos and the ICM. Future work will also explore in more detail the  pressure-anisotropy-driven instabilities of the slow-magnetosonic and CR-entropy modes (Section \ref{sec:slow}). We also plan to extend the CR--Braginskii MHD fluid model to collisionless models of the thermal plasma. 

Finally, we note the caveat that the dominant transport process of cosmic rays through galaxies and clusters remains uncertain (e.g., \citealt{amato_blasi_18}). For example, even cosmic rays that are not strongly coupled to the thermal plasma (i.e. not locked to the Alfv\'en frame, as a result of a low pitch-angle scattering rate, e.g. due to wave damping) may actually not be diffusing under certain conditions, but instead streaming at super-Alfv\'enic speeds (\citealt{skilling71}; \citealt{wiener2013}). The development of more accurate fluid models of cosmic rays is therefore a high priority.

\section*{Acknowledgements}
We thank R. Farber, Y. Jiang, M. Kunz, S. P. Oh, C. Pfrommer, M. Ruszkowski, A. Spitkovsky, J. Stone, I. Zhuravleva \& E. Zweibel for enlightening discussions. This research was supported in part by the Heising-Simons Foundation, the Simons Foundation, and National Science Foundation Grant No. NSF PHY-1748958, NSF grant AST-1715070,  and a Simons Investigator award from the Simons Foundation.  PK would like to thank the Kavli Institute for Theoretical Physics for their hospitality and support offered via the Graduate Fellowship Program. Support for J.S. was provided by Rutherford Discovery Fellowship RDF-U001804 and Marsden Fund grant UOO1727, which are managed through the Royal Society Te Ap\={a}rangi.

\bibliographystyle{mnras}

\bibliography{crbrag}


\appendix


\section{Acoustic Instability in Two-Fluid Plasma} \label{app:2F}
 
   \begin{figure}
  \centering
    \includegraphics[width=0.45 \textwidth]{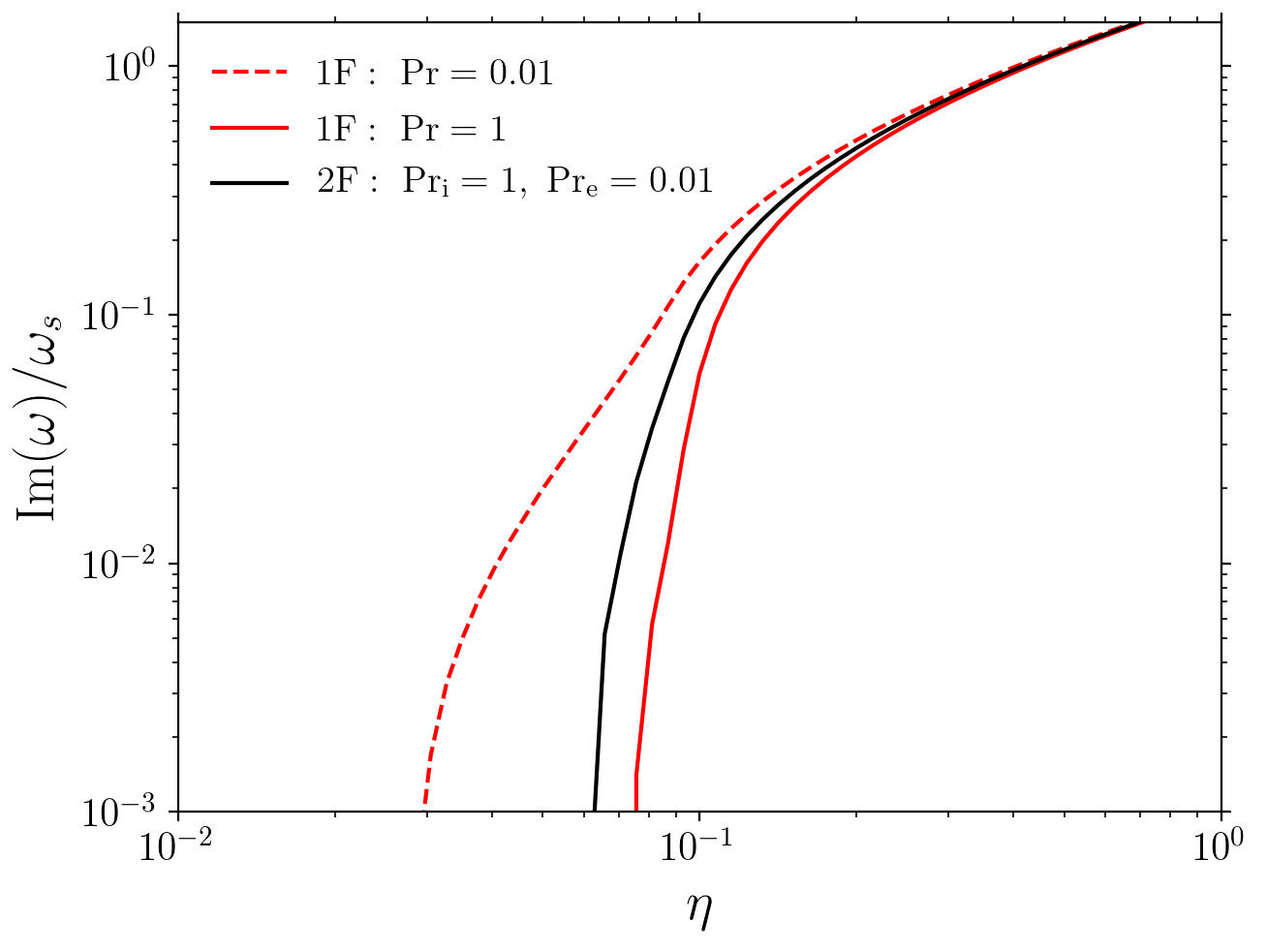}
 \caption{Comparison of the maximum sound-wave growth rates for the single-fluid (1F) and two-fluid (2F) ion-electron plasma as a function of $\eta$, for $\beta = 100$ and $\omega_d = 0$ (no CR diffusion). This figure is analogous to Figure \ref{fig:fast_fastest_eta}. We find that our results are not significantly affected by the extension to two entropy equations and heat fluxes. The two-fluid growth rate for $\nub / \chi_i = 1$ and $\nub / \chi_e = 0.01$ (black line) is essentially in between the 1-fluid prediction with ${\rm Pr} = 1$ and  ${\rm Pr} = 0.01$ (red lines). \label{fig:twoF} }
\end{figure}

In Section \ref{sec:params} we pointed out that the ion-electron temperature equilibration timescale is longer in the regime of interest than the relevant sound timescale. Using a single-fluid approach with heat flow carried by the electrons is then not correct. Instead, separate entropy equations and heat fluxes should be used for each species. In the main text, we only considered a single thermal fluid and a single heat flux for simplicity (with varying thermal Prandtl number), and we demonstrated that our results do not depend strongly on the value of the chosen thermal diffusivity. Here we show that our conclusions do not change in a more accurate two-fluid model, when separate electron and ion pressure equations are included.

In the two-fluid model we consider, the continuity, induction and CR pressure equations (\ref{eq:cont}, \ref{eq:induction} and \ref{eq:pc}) remain unchanged (we assume quasi-neutrality, $n_{\rm i} = n_{\rm e}$). The remaining equations that we need to solve are the momentum equation, and the ion and electron entropy equations:
\begin{equation} \label{eq:mom_2F}
\rho \frac{d \bvrm}{d t} = - \bm{\nabla} (p_i + p_e + p_c + \frac{B^2}{ 8 \pi}) + \frac{\bm{B \cdot \nabla B}}{4 \pi} +  \bm{\nabla \cdot } \big( \bm{\hat{b} \hat{b}} \Delta p \big) - \frac{1}{3} \bm{\nabla} \Delta p,
\end{equation}
\begin{gather}
\begin{aligned}\label{eq:pi}
    \frac{1}{\gamma -1} \frac{d p_i}{d t} & = - \frac{\gamma}{\gamma -1} p_i \bm{\nabla \cdot \vrm} - \alpha_1 \bm{\vadp \cdot \nabla}p_c  - \alpha_2 \bm{\nabla \cdot } \Big( \bm{\Pi \cdot \vrm} \Big) \\ & - \bm{\nabla \cdot Q_i} -  \frac{p_i - p_e}{\tau_{\rm eq}},
\end{aligned}
\end{gather}

\begin{gather}
\begin{aligned}\label{eq:pe}
    \frac{1}{\gamma -1} \frac{d p_e}{d t} & = -\frac{\gamma}{\gamma -1} p_e \bm{\nabla \cdot \vrm} - (1-\alpha_1) \bm{\vadp \cdot \nabla}p_c  \\ & - (1- \alpha_2) \bm{\nabla \cdot } \Big( \bm{\Pi \cdot \vrm} \Big)  - \bm{\nabla \cdot Q_e} +  \frac{p_i - p_e}{\tau_{\rm eq}}.
\end{aligned}
\end{gather}
$p_i$ and $p_e$ are the ion and electron pressures, respectively. $\tau_{\rm eq}$ is the timescale over which the electrons and ions come into thermal equilibrium. This timescale is long compared to the electron-electron and ion-ion collision times, $\tau_{\rm ee} \sim \sqrt{m_{\rm e} / m_{\rm i}} \tau_{\rm ii} \sim (m_{\rm e}/m_{\rm i}) \tau_{\rm  eq}$. The smallness of the equilibration term in a weakly collisional plasma is what motivates the 2-fluid model and so we will drop the terms $\propto \tau_{\rm  eq}^{-1}$ in equations \ref{eq:pi} and \ref{eq:pe} in this section. $\alpha_1$ and $\alpha_2$ are parameters which set how much of the CR and viscous heating goes into the ions vs. electrons. Viscous heating does not enter in our analysis to linear order, and so $\alpha_2$ can be ignored. We choose $\alpha_1 = 0.5$, but our results do not depend on it, as the instability is ultimately not driven by CR heating at high $\beta$. In \eqref{eq:pi} and \eqref{eq:pe}, $\bm{Q_i} = - n_{\rm i} k_{\rm B} \chi_i \bm{\hat{b}\hat{b}\cdot \nabla}T_i$ is the ion heat flux ($\chi_i$ is the ion thermal diffusivity) and $\bm{Q_e} = - n_{\rm e} k_{\rm B} \chi_e \bm{\hat{b}\hat{b}\cdot \nabla}T_e$ is the electron heat flux ($\chi_e$ is the electron thermal diffusivity).  

The linearised versions of \eqref{eq:pi} and \eqref{eq:pe} are:
\begin{equation} \label{eq:pi_pert}
    \omega \frac{\delta p_i}{p_i} = \gamma \bm{k \cdot \vrm} + 2 \alpha_1 (\gamma-1) \omega_a \eta \frac{\delta p_c}{p_c} - i (\gamma -1) \omega_{\rm cond,i} \Big( \frac{\delta p_i}{p_i} - \frac{\delta \rho}{\rho} \Big),
\end{equation}

\begin{equation} \label{eq:pe_pert}
    \omega \frac{\delta p_e}{p_e} = \gamma \bm{k \cdot \vrm} + 2(1-\alpha_1)(\gamma-1) \omega_a \eta \frac{\delta p_c}{p_c} - i (\gamma -1) \omega_{\rm cond,e} \Big( \frac{\delta p_e}{p_e} - \frac{\delta \rho}{\rho} \Big),
\end{equation}
where we defined the ion and electron thermal diffusion frequencies, $\omega_{\rm cond,i/e} = \chi_{i/e} (\bm{\hat{b}\cdot k})^2$. 

We assume an equilibrium with $p_i = p_e$. As in Section \ref{sec:1D}, we can derive a 1D dispersion relation for the two-fluid acoustic instability (again assuming high $\beta$, $\omega_s \gg \omega_a$):
\begin{gather}
\begin{aligned}\label{eq:1D_disp_2F}
    0 = & \omega^2 - \frac{\omega_s^2}{2 \gamma} \Big( \frac{\gamma \omega + i(\gamma - 1) \omega_{\rm cond,i}}{\omega + i (\gamma-1) \omega_{\rm cond,i}}  + \frac{\gamma \omega + i(\gamma - 1) \omega_{\rm cond,e}}{\omega + i (\gamma-1) \omega_{\rm cond,e}}   \Big) \\ & + \frac{4}{3}i \omega \omega_{\rm B} - \eta \frac{\omega_s^2}{\gamma} \Big( \frac{4}{3}+ \frac{8}{3}i \frac{\omega_{\rm B}}{\omega_a} \Big) \Big( 1 + i \frac{\omega_d}{\omega}  \Big)^{-1}.
\end{aligned}
\end{gather}
Since electron conduction is rapid, we can consider the regime $\omega_{\rm cond, e}\gg \omega_s$, such that the electrons are essentially isothermal. The above then simplifies to:
\begin{gather}
\begin{aligned}\label{eq:1D_disp_2F_faste}
    0 = & \omega^2 - \frac{\omega_s^2}{2 \gamma} \Big( \frac{\gamma \omega + i(\gamma - 1) \omega_{\rm cond,i}}{\omega + i (\gamma-1) \omega_{\rm cond,i}}  + 1  \Big)  + \frac{4}{3}i \omega \omega_{\rm B}  \\ & - \eta \frac{\omega_s^2}{\gamma} \Big( \frac{4}{3}+ \frac{8}{3}i \frac{\omega_{\rm B}}{\omega_a} \Big) \Big( 1 + i \frac{\omega_d}{\omega}  \Big)^{-1}
\end{aligned}
\end{gather}
Note that \eqref{eq:1D_disp_2F} and \eqref{eq:1D_disp_2F_faste} are very similar to the dispersion relation in \eqref{eq:1d_complete}. As a result, we find that our results are not significantly affected by the extension to two entropy equations and heat fluxes. This is confirmed in Figure \ref{fig:twoF}, where we see that the two-fluid fast magnetosonic growth rate for $\nub / \chi_i = 1$ and $\nub / \chi_e = 0.01$ is essentially in between the single-fluid prediction with ${\rm Pr} = 1$ and  ${\rm Pr} = 0.01$.

\bsp	
\label{lastpage}
\end{document}